\documentclass[pra,showpacs]{revtex4}
\usepackage{epsfig}
\usepackage{amssymb}


\begin{document}

\title{Dynamic correlation functions in one-dimensional quasi-condensates}

\author{D.L. Luxat}

\email{luxat@physics.utoronto.ca}

\author{A. Griffin}

\email{griffin@physics.utoronto.ca}

\affiliation{Department of Physics, University of Toronto, Toronto, Ontario, Canada
M5S 1A7}

\date{\today}


\begin{abstract}
We calculate the static and dynamic single-particle correlation functions
in one-dimensional (1D) trapped Bose gases and discuss experimental
measurements that can directly probe such correlation functions. Using
a quantized hydrodynamic theory for the low energy excitations, we calculate both the static and
dynamic single-particle correlation functions for a 1D Bose gas that is a phase-fluctuating quasi-condensate. For the static (equal-time) correlation
function, our approximations and results are equivalent to those of
Petrov, Shlyapnikov and Walraven. The Fourier transform of the static single-particle correlation
function gives the momentum distribution, which can be measured using Doppler-sensitive Bragg scattering experiments on a highly elongated Bose gas. We show how a two-photon Raman out-coupling experiment can measure the characteristic features of the dynamic or time-dependent
single-particle correlation function of a 1D Bose quasi-condensate.
\end{abstract}

\maketitle
\section{\label{sec:Introduction}Introduction}

In low-dimensional uniform Bose gases, the single-particle correlations
are fundamentally different from a three-dimensional gas, dramatically affecting the nature of the Bose condensation phase transition \cite{popov-collective-excitations}. G\"{o}rlitz et.~al.~\cite{gorlitz-prl-87-130402} have created one- and two-dimensional \emph{trapped} Bose gases by making the level spacing of the confining potential (along one or two directions) larger than the mean-field interaction energy. Recently, several experimental and theoretical studies have highlighted the manner in which the unique correlations of a uniform low-dimensional Bose gas are modified in a trapped Bose gas. With the increasing interest in BEC matter wave interferometry and atom lasers, it is clearly important to have a greater understanding of one-dimensional (1D) trapped Bose gases.

In uniform systems, it is well known that the behavior of Bose gases
is quite different in one- or two-dimensions compared to three-dimensions.
In 3D, below a critical temperature $T_{BEC}$, the
long-range correlations of the phase are manifested in the asymptotic
properties of the single-particle correlation function, where $\lim _{\left|\mathbf{r}\right|\rightarrow \infty }\left\langle \hat{\psi }^{\dagger }(\mathbf{r})\hat{\psi }(\mathbf{0})\right\rangle \sim n_{0}\neq 0$
indicates long-range order in the system. Above $T_{BEC}$, this single-particle
correlation function decays exponentially for large distances, characteristic
of normal systems. In systems of reduced dimensionality, long
wavelength fluctuations of the phase of the Bose order parameter lead to a new state intermediate between a Bose-condensed and non-Bose-condensed system. The system
forms a quasi-condensate, in which the phase of the order parameter is only coherent over a finite length scale. The system may be thought of as Bose condensate sub-systems which are well defined locally but are not correlated with each other \cite{popov-collective-excitations}.

The presence of a quasi-condensate causes the equal-time single-particle correlation
function to spatially decay with a characteristic power law, $\lim _{\left|\mathbf{r}\right|\rightarrow \infty }\left\langle \hat{\psi }^{\dagger }(\mathbf{r})\hat{\psi }(\mathbf{0})\right\rangle \sim 1/\left|\mathbf{r}\right|^{\alpha }$,
where the power law parameter $\alpha $ is characteristic of the
system. In a uniform two-dimensional (2D) Bose gas, a Bose condensate can be shown
to only occur at $T=0$. At low but finite temperatures, however, a 2D Bose gas can form a quasi-condensate. In a one-dimensional (1D) uniform Bose gas, true Bose
condensation never occurs, but at $T=0$ a quasi-condensate is possible.
The central point is that the characteristic features of 1D and 2D
quasi-condensates are associated with long wavelength, low frequency
(i.e., long distances and times) phase fluctuations of the Bose order parameter.

In a \emph{trapped} Bose gas, the effect of the long wavelength phase fluctuations
is complicated by the finite size of the system. While a quasi-condensate
is still locally coherent, the size of the system will play a crucial role. Recent theoretical work on trapped Bose gases has focussed on the
static correlation properties of trapped 1D and 2D Bose gases
as a measure of the long-range phase coherence of the Bose gas. Early work by
Ho and Ma \cite{ho-jltp-115-61} considered primarily the zero-temperature
(vacuum) phase fluctuations. They showed
that the strong phase fluctuations were restricted by the finite size of the system and as a result the system could
form a Bose condensate even at $T=0$. Since the low lying collective
modes have finite energy in a trapped Bose gas, at sufficiently low
temperatures the thermal fluctuations are frozen and the $T=0$
analysis still applies. The role of thermal fluctuations in determining
the phase coherence of the trapped Bose gas has been recently discussed
in considerable detail by Shlyapnikov and coworkers \cite{petrov-prl-84-2551,petrov-prl-85-3745,petrov-prl-87-050404,shlyapnikov-cargese-2-407}.
They demonstrated \cite{petrov-prl-85-3745} that below a certain
temperature $T_{\phi }$, the long wavelength phase fluctuations in
a trapped 1D Bose gas are effectively frozen. As a result, the quasi-condensate of the uniform gas becomes a true Bose condensate in the trapped gas for $T<T_{\phi}$. This work was later extended to deal with the effect of phase fluctuations in a highly elongated (cigar-shaped)
3D Bose gas \cite{petrov-prl-87-050404}, which was shown to behave as a quasi-1D Bose gas. It was found that the long wavelength phase fluctuations above a certain temperature $T_{\phi }^{3D}$ would still destroy long-range order, leading to a quasi-condensate.

The role of phase fluctuations in an elongated 3D Bose gas
has been recently explored by Dettmer et.~al.~\cite{dettmer-prl-87-160406},
where evidence was found for a phase-fluctuating quasi-condensate by observation of characteristic interference effects in the freely expanding gas. This kind of real space interferometry experiment is a measure of the equal-time single-particle correlations of a Bose gas. However, the theoretical analysis required to evaluate the effect of phase fluctuations is quite involved and somewhat indirect \cite{dettmer-prl-87-160406}. The Doppler-sensitive Bragg scattering experiments recently performed in Orsay \cite{gerbier-cond-mat-0210206} give an alternative and perhaps a more direct probe of the single-particle correlations, albeit in momentum space. 

Such Bragg scattering techniques have been used to measure the excitations in a trapped 3D Bose gas \cite{stenger-prl-82-4569,stamper-kurn-prl-83-2876,ketterle-les-houches-1999,vogels-prl-88-060402,steinhauer-prl-88-120407}.
Out-coupling (or tunneling) experiments discussed in Refs.~\cite{japha-prl-82-1079,choi-pra-61-063606,luxat-pra-65-043618}
also show how one can obtain information about the dynamic single-particle
correlation functions of a trapped Bose gas. There has been some related theoretical work on the absorption line shape of a uniform 1D Bose gas \cite{yip-prl-87-130401}.
This type of experiment is also similar to tunneling between quantum wires used to probe the single-particle correlations in 1D electron systems \cite{kang-nature-403-59,auslaender-science-295-825}.

We restrict ourselves to the trapped 1D mean-field
Bose gas, as defined in Refs.~\cite{petrov-prl-85-3745,shlyapnikov-cargese-2-407,menotti-pra-66-043610}.
In Section \ref{sec:Correlation-functions}, we summarize the calculation of low energy collective modes responsible
for the long wavelength phase fluctuations using a quantized hydrodynamic
theory \cite{stringari-prl-77-2360,wu-pra-54-4204}. In Section \ref{sec:Static-correlation-function}, we use these
results to calculate the static correlation function for a quasi-condensate, following the work of Shlyapnikov and coworkers ~\cite{petrov-prl-85-3745,petrov-prl-87-050404,shlyapnikov-cargese-2-407}. The results for the static correlation function of a quasi-condensate
are compared with that of a true Bose condensate in the 1D trap at $T=0$. We then
calculate the momentum distribution of atoms both in quasi-condensates
and true Bose condensates, since this is the key quantity measured
in Doppler-sensitive Bragg spectroscopy. This discussion of results for the static correlation function sets the stage for our generalization to 1D dynamic correlation functions in Section \ref{sec:Dynamic-correlation-function}. In Section \ref{sec:Out-coupling-experiment}, we show how this dynamic correlation function could be measured using a two-photon Raman scattering (out-coupling) experiment.
\section{\label{sec:Correlation-functions}Correlation functions}

The two-photon optical Bragg scattering cross-section can be shown to be proportional to \cite{ketterle-les-houches-1999,zambelli-pra-61-063608}
\begin{equation}
[S(\mathbf{q},\omega)-S(-\mathbf{q},-\omega)],
\label{eq:bragg-scattering-cross-section}
\end{equation}
where the dynamic structure factor is defined by
\begin{equation}
S(\mathbf{q},\omega)=\frac{1}{2\pi} \int_{-\infty}^{\infty}dte^{i\omega t}\int d\mathbf{r} \int d\mathbf{r}^{\prime} e^{-i\mathbf{q} \cdot (\mathbf{r}-\mathbf{r}^{\prime})} \left\langle \hat{n}(\mathbf{r},t)\hat{n}(\mathbf{r}^{\prime},0) \right \rangle ,
\label{eq:dynamic-structure-factor-defined}
\end{equation}
with $\hat{n}(\mathbf{r}) \equiv \hat{\psi}^{\dagger}(\mathbf{r})\hat{\psi}(\mathbf{r})$ being the local density operator. Quite generally, one can prove from detailed balancing that  $S(-\mathbf{q},-\omega)=e^{-\beta \omega} S(\mathbf{q},\omega)$. Even for a trapped gas, we can express the quantum field operators in the momentum representation
\begin{equation}
\hat{\psi}(\mathbf{r})=\int \frac{d\mathbf{k}}{(2\pi)^{3/2}} e^{i\mathbf{k} \cdot \mathbf{r}} \hat{\psi}(\mathbf{k}),
\label{eq:field-operator-in-momentum-representation}
\end{equation}
in which case (\ref{eq:dynamic-structure-factor-defined}) reduces to
\begin{equation}
S(\mathbf{q},\omega)=\frac{1}{2\pi} \int \frac{d\mathbf{k}d\mathbf{k}^{\prime}}{(2\pi)^{3}} \int_{-\infty}^{\infty} dte^{i\omega t} \left\langle \hat{\psi}^{\dagger}(\mathbf{k},t) \hat{\psi}(\mathbf{k}+\mathbf{q},t) \hat{\psi}^{\dagger}(\mathbf{k}^{\prime}) \hat{\psi}(\mathbf{k}^{\prime}-\mathbf{q}) \right\rangle.
\label{eq:dynamic-structure-factor-momentum-representation}
\end{equation}
As we review in an Appendix, within the impulse approximation (IA), one can reduce Eq.~(\ref{eq:dynamic-structure-factor-momentum-representation}) to the simple looking result
\begin{equation}
S(\mathbf{q},\omega)=\int \frac{d\mathbf{k}}{(2\pi)^{3}} n(\mathbf{k})\delta(\omega - q^{2}/2m - \mathbf{q} \cdot \mathbf{k}/m),
\label{eq:IA-dynamic-structure-factor}
\end{equation}
where $n(\mathbf{k}) \equiv \langle \hat{\psi}^{\dagger}(\mathbf{k}) \hat{\psi}(\mathbf{k}) \rangle$ is the thermally averaged equilibrium momentum distribution of atoms in a fully interacting trapped gas. This result also shows that the energy transfer in the IA is very large and hence the second term in Eq.~(\ref{eq:bragg-scattering-cross-section}) can be neglected ($\omega \gg k_{B}T$).

Thus one sees how Doppler-sensitive Bragg scattering gives information about the atomic momentum distribution $n(\mathbf{k})$. In turn, this momentum distribution is the Fourier transform of a single-particle correlation function, which we now discuss for a 1D Bose gas. We consider a Bose gas in the mean-field regime which is trapped in a highly elongated cigar-shaped trap. In this
1D geometry, the confinement along the radial direction is so tight
that the trap frequency along the radial direction $\omega _{\rho }$
is much greater than the mean-field interaction energy. At the low temperatures of interest,
the dynamics of the atoms in the radial direction are essentially
{}``frozen'', with all the atoms occupying the ground-state of the harmonic
trap in the radial direction
\begin{equation}
\phi _{\rho }(x,y)=\sqrt{\frac{1}{\pi l_{\rho }^{2}}}e^{-(x^{2}+y^{2})/2l_{\rho }^{2}}.
\label{eq:radial-wavefunction}
\end{equation}
Here the extension of the wavefunction in the radial direction is
given by the harmonic oscillator length $l_{\rho }\equiv \sqrt{\hbar /m\omega _{\rho }}$.

In the mean-field regime, $l_{\rho}$ is much greater than the radius
of the interatomic potential $R_{e}$. The scattering of atoms in
this 1D system is thus still three-dimensional. According to Ref.~\cite{olshanii-prl-81-938},
the effective coupling constant in this 1D system is
\begin{equation}
g_{1D}=\frac{2\hbar ^{2}}{m}\frac{a}{l_{\rho }(l_{\rho }-Ca)},
\label{eq:olshanii-g1D}
\end{equation}
where $a$ is the s-wave scattering length and $C$ is a numerical
constant on the order of $1$. The term $Ca$ in Eq.~(\ref{eq:olshanii-g1D})
is negligible for $l_{\rho }\gg R_{e}$. In this limit, the expression
for $g_{1D}$ is the same as that obtained by averaging over the
radial wavefunction (\ref{eq:radial-wavefunction}) \cite{olshanii-prl-81-938,petrov-prl-85-3745},
\begin{eqnarray}
g_{1D} & \equiv & g \int_{-\infty}^{\infty}dx dy \phi_{\rho}(x,y)^{4}\nonumber \\
& = & \frac{2\hbar^{2}a}{m l_{\rho}^{2}}.
\label{eq:g1D}
\end{eqnarray}
We use this expression in the rest of this paper.

Since the trap frequency along the radial direction is much greater
than the mean-field interaction energy, in the mean-field regime we
can construct a 1D Gross-Pitaevskii equation for the axial direction
only. Within the Thomas-Fermi regime, the 1D Gross-Pitaevskii equation
leads to a density distribution in the axial direction given by
\begin{equation}
n_{0}(z)=\frac{1}{g_{1D}}\left(\mu _{0}-\frac{1}{2}m\omega _{z}^{2}z^{2}\right)\theta \left(\mu _{0}-\frac{1}{2}m\omega _{z}^{2}z^{2}\right),
\label{eq:axial-condensate-density}
\end{equation}
where the chemical potential is
\begin{equation}
\mu _{0}=\hbar \omega _{z}\left(\frac{3}{4\sqrt{2}}\frac{g_{1D}}{\hbar \omega _{z}l_{z}}N\right)^{2/3}.
\label{eq:condensate-chemical-potential}
\end{equation}
The Thomas-Fermi length of the atom cloud in the \emph{axial direction} is
then defined by
\begin{equation}
Z_{0}\equiv \sqrt{\frac{2\mu _{0}}{m\omega _{z}^{2}}}.
\label{eq:condensate-axial-length}
\end{equation}
The number of atoms is given by $N$ and $l_{z}\equiv \sqrt{\hbar /m\omega _{z}}$
is the harmonic oscillator length in the axial direction.

The theory of the correlation properties of a phase fluctuating Bose
gas builds upon this zero temperature theory of the static condensate
density. The central idea is that the time-dependent single-particle
correlation function
\begin{equation}
C_{\psi ^{\dagger }\psi }(\mathbf{r},\mathbf{r}^{\prime };t)\equiv \left\langle \hat{\psi }^{\dagger }(\mathbf{r},t)\hat{\psi }(\mathbf{r}^{\prime })\right\rangle 
\label{eq:single-particle-correlation-function-definition}
\end{equation}
can be calculated using the $T=0$ low energy collective modes calculated
as fluctuations about the static condensate density, even when we
are dealing with a quasi-condensate with large phase fluctuations at finite temperature. In calculating the single-particle correlation function in Eq.~(\ref{eq:single-particle-correlation-function-definition}), we write the Bose field operator as $\hat{\psi }(\mathbf{r},t)=\sqrt{\hat{n}(\mathbf{r},t)}e^{i\hat{\phi }(\mathbf{r},t)}$
using the density ($\hat{n}$) and phase ($\hat{\phi }$) variables. While the density fluctuations are small, the phase fluctuations can be large in 1D and 2D Bose gases. Within this approximation, only the phase fluctuation dynamics are considered and then the single-particle correlation function is given by
\begin{equation}
C_{\psi ^{\dagger }\psi }(\mathbf{r},\mathbf{r}^{\prime };t)\simeq \sqrt{n_{0}(\mathbf{r})n_{0}(\mathbf{r}^{\prime })}\left\langle e^{-i\left[\hat{\phi }(\mathbf{r},t)-\hat{\phi }(\mathbf{r}^{\prime })\right]}\right\rangle.
\label{eq:phase-fluctuating-single-particle-correlation-function}
\end{equation}
It is adequate to use the static condensate density $n_{0}(\mathbf{r})$ at $T=0$ given in Eq.~(\ref{eq:phase-fluctuating-single-particle-correlation-function}), since this is unaffected by phase fluctuations at low temperatures.

To evaluate the single-particle correlation function in
Eq.~(\ref{eq:phase-fluctuating-single-particle-correlation-function}),
we need the normal modes corresponding to low energy phase fluctuations. We use the quantized hydrodynamic theory developed by Wu and Griffin
\cite{wu-pra-54-4204}. The fluctuations in the density and phase about their equilibrium values are given by
\begin{equation}
\hat{n}(z,t)=n_{0}(z)+\delta \hat{n}(z,t)
\label{eq:density-operator--quantized-hydrodynamics}
\end{equation}
and
\begin{equation}
\hat{\phi }(z,t)=\phi _{0}(z)+\delta \hat{\phi }(z,t).
\label{eq:phase-operator--quantized-hydrodynamics}
\end{equation}
By starting with the Gross-Pitaevskii Hamiltonian, we may develop a quantized hydrodynamic Hamiltonian describing the low frequency excitations at $T=0$. Working to lowest order in the fluctuations
and making the Thomas-Fermi approximation, one obtains the effective hydrodynamic Hamiltonian \cite{wu-pra-54-4204}
\begin{equation}
\hat{H}=H_{0}+\int _{-\infty }^{\infty }dz\left[mn_{0}(z)\delta \hat{v}_{z}^{2}(z)+g_{1D}\delta \hat{n}^{2}(z)\right],
\label{eq:hydrodynamic-Hamiltonian}
\end{equation}
in which the fluctuations of the condensate phase and density are treated up
to quadratic order. The velocity fluctuation operator $\delta \hat{v}_{z}$
is related to the phase fluctuation operator in the usual way,
\begin{equation}
\delta \hat{v}_{z}(z,t)\equiv \frac{\hbar }{m}\frac{d}{dz}\delta \hat{\phi }(z,t).
\label{eq:velocity-fluctuation-operator--definition}
\end{equation}
The quadratic Hamiltonian in Eq.~(\ref{eq:hydrodynamic-Hamiltonian})
has the same form as the renormalized 1D hydrodynamic Hamiltonian
derived by Haldane \cite{haldane-prl-47-1840}, which has been used in recent studies on 1D Bose gases \cite{gangardt-lanl-0207338}.

Since the hydrodynamic Hamiltonian in Eq.~(\ref{eq:hydrodynamic-Hamiltonian})
is quadratic in the density and phase fluctuation operators, one can diagonalize it using the canonical transformations \cite{wu-pra-54-4204}
\begin{equation}
\delta \hat{n}(z,t)=\sum _{j}\left[A_{j}\psi _{j}(z)e^{-i\omega _{j}t}\hat{\alpha }_{j}+A_{j}^{*}\psi _{j}^{*}(z)e^{i\omega _{j}t}\hat{\alpha }_{j}^{\dagger }\right]\label{eq:density-fluctuation-operator--canonical-transformation}
\end{equation}
and
\begin{equation}
\delta \hat{\phi }(z,t)=\sum _{j}\left[B_{j}\psi _{j}(z)e^{-i\omega _{j}t}\hat{\alpha }_{j}+B_{j}^{*}\psi _{j}^{*}(z)e^{i\omega _{j}t}\hat{\alpha }_{j}^{\dagger }\right].
\label{eq:phase-fluctuation-operator--canonical-transformation}
\end{equation}
This form of the canonical transformations
ensures the Hermiticity of the density and phase operators. The operators
$\hat{\alpha }_{j}$ and $\hat{\alpha }_{j}^{\dagger }$ annihilate
and create excitations with energy
$\hbar \omega _{j}$, and satisfy the commutation relations $[\hat{\alpha }_{j},\hat{\alpha }_{j^{\prime }}]=[\hat{\alpha }_{j}^{\dagger },\hat{\alpha }_{j^{\prime }}^{\dagger }]=0$
and $[\hat{\alpha }_{j},\hat{\alpha }_{j^{\prime }}^{\dagger }]=\delta _{j,j^{\prime }}$.
The mode expansion coefficients $A_{j}$ and $B_{j}$ are to be determined
along with the mode eigenfunction $\psi _{j}(z)$ and the excitation
energy $\hbar \omega _{j}$ \cite{wu-pra-54-4204}. The mode eigenfunctions
are assumed to form a complete, orthonormal basis. From the equal-time
commutation relation $[\delta \hat{n}(z),\delta \hat{\phi }(z^{\prime })]=i\delta (z-z^{\prime })$,
the coefficients $A_{j}$ and $B_{j}$ may be shown to satisfy the
relation $A_{j}^{*}B_{j}=-i/2$.

Using the Hamiltonian in Eq.~(\ref{eq:hydrodynamic-Hamiltonian}), the density and phase fluctuation operators satisfy the Heisenberg equations of motion
\begin{equation}
\frac{\partial \delta \hat{n}(z,t)}{\partial t}=-\frac{\partial }{\partial z}\left[n_{0}(z)\delta \hat{v}_{z}(z,t)\right]
\label{eq:density-fluctuation-operator--Heisenberg-equation-of-motion}
\end{equation}
and
\begin{equation}
\frac{\partial \delta \hat{v}_{z}(z,t)}{\partial t}=-\frac{\partial }{\partial z}\left[g_{1D}\delta \hat{n}(z,t)\right].
\label{eq:velocity-fluctuation-operator--Heisenberg-equation-of-motion}
\end{equation}
These two equations may be combined to give the decoupled equations
\begin{equation}
\frac{\partial ^{2}\delta \hat{n}(z,t)}{\partial t^{2}}=\frac{\partial }{\partial z}\left[\frac{g_{1D}}{m}n_{0}(z)\frac{\partial }{\partial z}\delta \hat{n}(z,t)\right]
\label{eq:density-fluctuation-operator--PDE}
\end{equation}
and
\begin{equation}
\frac{\partial ^{2}\delta \hat{\phi }(z,t)}{\partial t^{2}}=\frac{\partial }{\partial z}\left[\frac{g_{1D}}{m}n_{0}(z)\frac{\partial }{\partial z}\delta \hat{\phi }(z,t)\right].
\label{eq:phase-fluctuation-operator--PDE}
\end{equation}
Using Eq.~(\ref{eq:density-fluctuation-operator--Heisenberg-equation-of-motion})
or Eq.~(\ref{eq:velocity-fluctuation-operator--Heisenberg-equation-of-motion})
together with the relation $A_{j}^{*}B_{j}=-i/2$, the expansion coefficients $A_{j}$ and $B_{j}$ are given by
\begin{equation}
A_{j}=i\sqrt{\frac{\hbar \omega _{j}}{2g_{1D}}},\;\;\; B_{j}=\sqrt{\frac{g_{1D}}{2\hbar \omega _{j}}}.
\label{eq:Aj-Bj--canonical-transformation-expansion-coefficients}
\end{equation}

Using Eq.~(\ref{eq:phase-fluctuation-operator--PDE}),
the equation for the eigenfunctions $\psi_{j}$ is
\begin{equation}
\frac{g_{1D}}{m}\frac{\partial }{\partial z}\left[n_{0}(z)\frac{\partial }{\partial z}\psi _{j}(z)\right]+\omega _{j}^{2}\psi _{j}(z)=0,
\label{eq:canonical-transformation-eigenfunction--ODE}
\end{equation}
with the static density $n_{0}(z)$ given by Eq.~(\ref{eq:axial-condensate-density}). Eq.~(\ref{eq:canonical-transformation-eigenfunction--ODE})
then reduces to a Legendre equation, with the eigenfrequencies
given by $\omega _{j}=\omega _{z}\sqrt{j(j+1)/2}$ \cite{stringari-pra-58-2385}. The associated eigenfunctions $\psi _{j}$ are given by
\begin{equation}
\psi _{j}(z)=\sqrt{\frac{j+1/2}{Z_{0}}}P_{j}(z/Z_{0}),
\label{eq:canonical-transformation-eigenfunction--solution}
\end{equation}
where $P_{j}$ is the usual Legendre polynomial and $Z_{0}$ is defined
in Eq.~(\ref{eq:condensate-axial-length}). This expression
for $\psi _{j}$ is normalized to unity and the integer $j$
runs from $1$ to $\infty $. Using these eigenfunctions, the density and phase fluctuation operators in Eq.~(\ref{eq:density-fluctuation-operator--canonical-transformation})
and Eq.~(\ref{eq:phase-fluctuation-operator--canonical-transformation})
are
\begin{equation}
\delta \hat{n}(z,t) = \sum_{j=1}^{\infty} \left[i\sqrt{\frac{\hbar \omega_{j}}{2g_{1D}}}\sqrt{\frac{j+1/2}{Z_{0}}}P_{j}(z/Z_{0}) e^{-i\omega_{j}t}\hat{\alpha }_{j}+H.c.\right]
\label{eq:density-fluctuation-operator--canonical-transformation-solved}
\end{equation}
and
\begin{equation}
\delta \hat{\phi }(z,t)=\sum _{j=1}^{\infty }\left[\sqrt{\frac{g_{1D}}{2\hbar \omega_{j}}}\sqrt{\frac{j+1/2}{Z_{0}}}P_{j}(z/Z_{0})e^{-i\omega_{j}t}\hat{\alpha }_{j}+H.c.\right].
\label{eq:phase-fluctuation-operator--canonical-transformation-solved}
\end{equation}
Identical results are obtained in Refs.~\cite{petrov-prl-85-3745,khawaja-pra-66-013615} by directly solving the coupled Bogoliubov equations for the 1D Bose gas. We note that the
low energy modes of a highly elongated trap are very similar
to those of the 1D Bose gas considered here. In the former case, we have $\omega_{j}=\omega _{z}\sqrt{j(j+3)/4}$,
while the wavefunctions are Jacobi polynomials $P_{j}^{(1,1)}$ instead of
Legendre polynomials \cite{petrov-prl-87-050404}. 

As noted in Ref.~\cite{petrov-prl-85-3745}, for low temperatures, the density fluctuations described by $\langle \delta \hat{n}(z)\delta \hat{n}(z^{\prime })\rangle $
are small in comparison with the phase fluctuations $\langle \delta \hat{\phi }(z)\delta \hat{\phi }(z^{\prime })\rangle $. We can understand this from the fact that for reasonable
parameter values, we have $g_{1D}>\hbar \omega _{j}$ for the low-lying normal
modes, which in turn implies $A_{j}\ll B_{j}$. This justifies ignoring
density fluctuations in reducing the correlation function in Eq.~(\ref{eq:single-particle-correlation-function-definition})
to Eq.~(\ref{eq:phase-fluctuating-single-particle-correlation-function}).

Within this mean-field approximation for the phase fluctuations, the dynamic single-particle correlation function in Eq.~(\ref{eq:phase-fluctuating-single-particle-correlation-function})
is given by
\begin{eqnarray}
C_{\psi ^{\dagger }\psi }(\mathbf{r},\mathbf{r}^{\prime };t) & = & \phi _{\rho }(x,y)\phi _{\rho }(x^{\prime },y^{\prime })\sqrt{n_{0}(z)n_{0}(z^{\prime })}\nonumber \\
 &  & \times e^{-\frac{1}{2}\left\langle \left[\hat{\delta \phi }(z,t)-\delta \hat{\phi }(z^{\prime },0)\right]^{2}\right\rangle }.
\label{eq:phase-fluctuating-single-particle-correlation-function--mean-field-approximation}
\end{eqnarray}
Here we have used the standard results for functional integration \cite{popov-collective-excitations} involving a quadratic Hamiltonian in reducing $\langle \exp (-i[\delta \hat{\phi }(z,t)-\delta \hat{\phi }(z^{\prime },0)])\rangle =\exp (-\frac{1}{2} \langle [\delta \hat{\phi }(z,t)-\delta \hat{\phi }(z^{\prime },0)]^{2}\rangle )$. For notational
convenience, we define $F(z,z^{\prime };t)\equiv \langle [\delta \hat{\phi }(z,t)-\delta \hat{\phi}(z^{\prime },0)]^{2}\rangle $. In Section \ref{sec:Static-correlation-function}, we evaluate and review the properties of this single-particle correlation function $C_{\psi ^{\dagger }\psi }(\mathbf{r},\mathbf{r}^{\prime };t)$ in the static limit ($t=0$). Section IV considers the extension to non-zero times.
\section{\label{sec:Static-correlation-function}Static correlation function}

Up to now, only the \emph{static} correlation function has been investigated
in a trapped low-dimensional Bose gas. Dettmer et.~al.~\cite{dettmer-prl-87-160406}
have provided impressive interferometric evidence for phase fluctuations in a tight cigar-shaped 3D Bose gas. Doppler-sensitive Bragg scattering experiments provide an alternative probe of the static single-particle correlation function since they measure the momentum distribution of
atoms in the quasi-condensate \cite{gerbier-cond-mat-0210206}.
In such a Bragg scattering experiment on a quasi-1D
highly elongated 3D system, a large momentum kick $\mathbf{q}$ is given to the
atoms by a Bragg pulse directed along the long axis of the trap. The signal is proportional to the dynamic structure factor in the high momentum regime \cite{zambelli-pra-61-063608}, the IA expression given in Eq.~(\ref{eq:IA-dynamic-structure-factor}).

The results of the present section are equivalent to those of Shlyapnikov and coworkers \cite{petrov-prl-85-3745,petrov-prl-87-050404,shlyapnikov-cargese-2-407}. However, in previous studies, the results were used to discuss interference patterns of expanding gases \cite{petrov-prl-85-3745}. In this section, we work directly with the dynamic structure factor measured in Bragg scattering experiments. These results will be needed in our generalization in Section \ref{sec:Dynamic-correlation-function} to time-dependent single-particle correlation functions in 1D Bose gases.

As in the case of a highly elongated 3D Bose gas, the dynamic structure
factor is the experimental signal in a Doppler-sensitive Bragg scattering
experiment on a 1D Bose gas. For the two Bragg laser beams oriented
so that the momentum kick given to the atoms is along the long axial
direction, the dynamic structure factor in the impulse approximation given by Eq.~(\ref{eq:IA-dynamic-structure-factor}) is
\begin{equation}
S_{IA}(\mathbf{q}=q_{z}\hat{\mathbf{z}},\omega )=\int \frac{d\mathbf{k}}{(2\pi)^{3}} n(\mathbf{k})\delta (\omega -\hbar q_{z}^{2}/2m-\hbar q_{z}k_{z}/m).
\label{eq:IA-dynamic-structure-factor-specific-to-experiment}
\end{equation}
This may be reduced to
\begin{equation}
S_{IA}(\mathbf{q}=q_{z}\hat{\mathbf{z}},\omega )=\frac{1}{(2\pi)^{3}} \sqrt{ \frac{m}{2\hbar \omega_{r}}}\int _{-\infty }^{\infty }dk_{x}\int _{-\infty }^{\infty }dk_{y}n(k_{x},k_{y},k_{z}=\sqrt{m/2 \hbar \omega_{r}}(\omega -\omega _{r})),
\label{eq:IA-dynamic-structure-factor-specific-to-experiment-simplified}
\end{equation}
where we have introduced the recoil energy $\hbar\omega _{r}\equiv \hbar^{2} q_{z}^{2}/2m$.
This shows how the IA dynamic structure factor is broadened as a result of the Doppler effect, which gives a fairly direct measure of the width of the momentum distribution of the atoms. The standard approach to estimate the broadening of the dynamic structure factor is to fit the data to a Gaussian centred at the peak value $\omega =\omega_{r}$,
\begin{equation}
S_{IA}(\mathbf{q},\omega )=S_{IA}(\mathbf{q},\omega =\omega _{r})\exp \left[-\frac{(\omega -\omega _{r})^{2}}{2\Delta _{IA}^{2}}\right],
\label{eq:IA-dynamic-structure-factor-Gaussian-expansion}
\end{equation}
where $\Delta _{IA}$ is the Doppler width of the dynamic structure
factor, given by 
\begin{equation}
\Delta _{IA}^{2}=-\left[S_{IA}(\mathbf{q},\omega )/\partial _{\omega }^{2}S_{IA}(\mathbf{q},\omega )\right]\left|_{\omega =\omega _{r}}\right..
\label{eq:IA-Doppler-width}
\end{equation}
For further discussion, see Ref.~\cite{zambelli-pra-61-063608}.

We now turn to evaluating the dynamic structure factor in Eq.~(\ref{eq:IA-dynamic-structure-factor-specific-to-experiment-simplified}) in two different
regimes: a Bose-condensed and a quasi-condensate gas. We first calculate the momentum distribution. Since both regimes occur at relatively low temperatures, we ignore thermally-excited density fluctuations. However, since these do not significantly affect the transition from a Bose condensate to a quasi-condensate as the temperature increases above $T_{\phi}$ {[}see Eq.~(\ref{eq:Tphi}){]}, we are not neglecting any
key features that a Doppler-sensitive Bragg scattering experiment
probes. A more sophisticated theory must be used to take into account the density fluctuations that become increasingly important at higher temperatures (see, e.g., Refs.~\cite{andersen-prl-88-070407,khawaja-pra-66-013615}).

We first consider the momentum distribution for a true Bose condensate
at $T=0$. The static correlation function can be split into a static Bose condensate term and the density fluctuations about the condensed state
\begin{equation}
\left\langle \hat{\psi }^{\dagger }(\mathbf{r})\hat{\psi }(\mathbf{r}^{\prime })\right\rangle =\Phi _{0}^{*}(\mathbf{r})\Phi _{0}(\mathbf{r}^{\prime })+\left\langle \tilde{\psi }^{\dagger }(\mathbf{r})\tilde{\psi }(\mathbf{r}^{\prime })\right\rangle .
\label{eq:field-operator-decomposition-when-symmetry-broken}
\end{equation}
The density fluctuations about the static Bose condensate are small, and thus we ignore the second term in (\ref{eq:field-operator-decomposition-when-symmetry-broken}), as discussed above.
The momentum distribution of atoms in the Bose-condensed gas is then
\begin{equation}
n_{c}(\mathbf{k})=\int d\mathbf{r}\, e^{i\mathbf{k}\cdot \mathbf{r}}\Phi _{0}^{*}(\mathbf{r})\int d\mathbf{r}^{\prime }\, e^{-i\mathbf{k}\cdot \mathbf{r}^{\prime }}\Phi _{0}(\mathbf{r}^{\prime }).
\label{eq:momentum-distribution-of-Bose-condensate}
\end{equation}
As discussed in Section \ref{sec:Correlation-functions}, the order parameter of a 1D Bose-condensed
gas is given by the harmonic oscillator ground state wavefunction
{[}see Eq.~(\ref{eq:radial-wavefunction}){]} in the radial direction,
and a Thomas-Fermi order parameter along the axial direction. The momentum distribution then reduces to
\begin{eqnarray}
n_{c}(\mathbf{k}) & = & \int d\mathbf{r}\, e^{i\mathbf{k}\cdot \mathbf{r}}\phi _{\rho }(x,y)\sqrt{n_{0}(z)}\nonumber \\
 &  & \times \int d\mathbf{r}^{\prime }\, e^{-i\mathbf{k}\cdot \mathbf{r}^{\prime }}\phi _{\rho }(x^{\prime },y^{\prime })\sqrt{n_{0}(z^{\prime })},
\label{eq:momentum-distribution-of-quasi-1D-Bose-condensate}
\end{eqnarray}
where the static condensate density $n_{0}(z)$ is defined in (\ref{eq:axial-condensate-density})
\cite{petrov-prl-85-3745}. 

The integrals in (\ref{eq:momentum-distribution-of-quasi-1D-Bose-condensate})
may be evaluated analytically to give
\begin{equation}
n_{c}(k_{x},k_{y},k_{z})=4\pi ^{3}l_{\rho }^{2}\frac{\mu _{0}}{g_{1D}}e^{-l_{\rho }^{2}(k_{x}^{2}+k_{y}^{2})}\frac{J_{1}^{2}(Z_{0}k_{z})}{k_{z}^{2}},
\label{eq:momentum-distribution-of-quasi-1D-Bose-condensate--evaluated}
\end{equation}
where $J_{1}$ is the Bessel function of the first kind of order 1. Using Eq.~(\ref{eq:momentum-distribution-of-quasi-1D-Bose-condensate--evaluated})
for the momentum distribution of a Bose condensate, the IA dynamic structure factor in Eq.~(\ref{eq:IA-dynamic-structure-factor-specific-to-experiment-simplified})
for a Bose-condensed gas is given by
\begin{equation}
S_{IA}^{c}(\mathbf{q}=q_{z}\hat{\mathbf{z}},\omega )=\frac{\pi}{2} \frac{\mu _{0}}{g_{1D}}\sqrt{\frac{2 \hbar \omega _{r}}{m}}\frac{J_{1}^{2}(Z_{0}\sqrt{m/2 \hbar \omega _{r}}(\omega -\omega _{r}))}{(\omega -\omega _{r})^{2}}.
\label{eq:IA-dynamic-structure-factor-condensate}
\end{equation}

To determine the momentum distribution for a quasi-condensate,
we must first evaluate the single-particle correlation function in Eq.~(\ref{eq:phase-fluctuating-single-particle-correlation-function--mean-field-approximation})
in the static limit ($t=0$),
\begin{eqnarray}
C_{\psi ^{\dagger }\psi }(\mathbf{r},\mathbf{r}^{\prime };t=0) & = & \phi _{\rho }(x,y)\phi _{\rho }(x^{\prime },y^{\prime })\sqrt{n_{0}(z)n_{0}(z^{\prime })}\nonumber \\
 &  & \times e^{-\frac{1}{2}F_{s}(z,z^{\prime })},
\label{eq:static-single-particle-correlation-function--quasi-condensate}
\end{eqnarray}
where $F_{s}(z,z^{\prime })\equiv F(z,z^{\prime };t=0)$. Using the result in Eq.~(\ref{eq:phase-fluctuation-operator--canonical-transformation-solved})
for the phase fluctuation operator, one obtains \cite{petrov-prl-85-3745}
\begin{eqnarray}
F_{s}(z,z^{\prime }) & = & \frac{g_{1D}}{2Z_{0}}\sum _{j=1}^{\infty }\frac{j+1/2}{\hbar \omega_{j}}\nonumber \\
 &  & \times \left[P_{j}(z/Z_{0})-P_{j}(z^{\prime }/Z_{0})\right]^{2}\nonumber \\
 &  & \times \left[1+2N(\hbar \omega_{j})\right],
\label{eq:phase-argument--static}
\end{eqnarray}
where $N(\hbar \omega_{j})$ is the Bose distribution function. In approximating
this expression further, we make use of two facts. The vacuum fluctuations
are small \cite{petrov-prl-85-3745} and secondly, we are interested
in the finite temperature regime where the thermally induced low energy
phase fluctuations destroy the long-range order of the Bose condensate.
We may then use the quasi-classical approximation in which the temperature
is larger than the important low energy excitations. Replacing
$N(\hbar \omega_{j})$ by $k_{B}T/\hbar \omega_{j}$, Eq.~(\ref{eq:phase-argument--static})
reduces to
\begin{eqnarray}
F_{s}(z,z^{\prime }) & = & \frac{4}{3}\frac{T\mu _{0}}{T_{d}\hbar \omega _{z}}\sum _{j=1}^{\infty }\frac{2j+1}{j(j+1)}\nonumber \\
 &  & \times \left[P_{j}(z/Z_{0})-P_{j}(z^{\prime }/Z_{0})\right]^{2}.
\label{eq:thermal-phase-argument--static}
\end{eqnarray}
Here the $T=0$ excitation energies are $\omega_{j}=\omega _{z}\sqrt{j(j+1)/2}$ and following Ref.~\cite{petrov-prl-85-3745}, we have introduced the {}``degeneracy temperature'' $T_{d}\equiv N\hbar \omega _{z}$. The $j$-sum over the modes may be evaluated
exactly,
\begin{eqnarray}
F_{s}(z,z^{\prime }) & = & \frac{4}{3}\frac{T\mu _{0}}{T_{d}\hbar \omega _{z}}\nonumber \\
 &  & \times \left|\ln \left[\frac{(1-z/Z_{0})(1+z^{\prime }/Z_{0})}{(1+z/Z_{0})(1-z^{\prime }/Z_{0})}\right]\right|.
\label{eq:thermal-phase-argument--static--evaluated}
\end{eqnarray}
We note that Eq.~(\ref{eq:thermal-phase-argument--static--evaluated}) agrees precisely with that obtained in Ref.~\cite{petrov-prl-85-3745} by solving the coupled Bogoliubov
equations, as it should. 

From Eq.~(\ref{eq:thermal-phase-argument--static--evaluated}),
it follows that the temperature $T_{\phi }$ at which phase fluctuations
destroy the long-range phase coherence of the Bose condensate can be characterized by \cite{petrov-prl-85-3745}
\begin{equation}
T_{\phi}=\frac{T_{d}\hbar \omega _{z}}{\mu _{0}}.
\label{eq:Tphi}
\end{equation}
Working to lowest order in the separation $\left|z-z^{\prime }\right|$ in Eq.~(\ref{eq:thermal-phase-argument--static--evaluated}), the single-particle
correlation function can be shown to decay as $\exp (-|z-z^{\prime }|/l_{\phi })$
near the centre of the trap, where the correlation length is \cite{petrov-prl-85-3745}
\begin{equation}
l_{\phi}=Z_{0}\frac{T_{\phi }}{T}.
\label{eq:phase-coherence-length}
\end{equation}
A quasi-1D Bose gas in a highly elongated trap has a phase fluctuation temperature
given by \cite{petrov-prl-87-050404}
\begin{equation}
T_{\phi }^{3D}=\frac{15}{32}\frac{T_{d}\hbar \omega _{z}}{\mu _{0}}
\label{eq:Tphi3D}
\end{equation}
with the correlation length
\begin{equation}
l_{\phi }^{3D}=Z_{0}\frac{T_{\phi }^{3D}}{T}.
\label{eq:phase-coherence-length-3D}
\end{equation}
These results are essentially the same as those for trapped 1D Bose gases which we work with.

We can now use these results to calculate the momentum distribution for a quasi-condensate (denoted by the subscript $qc$),
\begin{equation}
n_{qc}(\mathbf{k})=C_{\psi^{\dagger}\psi}(-\mathbf{k},\mathbf{k})=\int d\mathbf{r}e^{i\mathbf{k}\cdot \mathbf{r}}\int d\mathbf{r}^{\prime }e^{-i\mathbf{k}\cdot \mathbf{r}^{\prime }}C_{\psi ^{\dagger }\psi }(\mathbf{r},\mathbf{r}^{\prime };t=0).
\label{eq:quasi-condensate-momentum-distribution}
\end{equation}
The integration over the radial coordinates may be done analytically, and we
find that the momentum distribution of particles in the radial
direction is a Gaussian,
\begin{eqnarray}
n_{qc}(k_{x},k_{y},k_{z}) & = & 4\pi l_{\rho }^{2}e^{-l_{\rho }^{2}(k_{x}^{2}+k_{y}^{2})}\int _{-\infty }^{\infty }dze^{ik_{z}z}\int _{-\infty }^{\infty }dz^{\prime }e^{-ik_{z}z^{\prime }}\nonumber \\
 &  & \times \sqrt{n_{0}(z)n_{0}(z^{\prime })}e^{-\frac{1}{2}F_{s}(z,z^{\prime })}.
\label{eq:quasi-condensate-momentum-distribution--more-explicit}
\end{eqnarray}
The dependence of $n_{qc}$ on the axial momentum $k_{z}$
is clearly complicated. Substituting the expressions for $n_{0}(z)$
in Eq.~(\ref{eq:axial-condensate-density}) and $F_{s}(z,z^{\prime })$
in Eq.~(\ref{eq:thermal-phase-argument--static--evaluated}) into
Eq.~(\ref{eq:quasi-condensate-momentum-distribution--more-explicit}),
the axial quasi-condensate momentum distribution must be evaluated numerically.
Substituting Eq.~(\ref{eq:quasi-condensate-momentum-distribution--more-explicit})
into Eq.~(\ref{eq:IA-dynamic-structure-factor-specific-to-experiment-simplified}),
the IA quasi-condensate dynamic structure factor is
\begin{equation}
S_{IA}^{qc}(\mathbf{q}=q_{z}\hat{\mathbf{z}},\omega )=\frac{1}{2\pi}\sqrt{\frac{m}{2 \hbar \omega_{r}}}n_{qc}(k_{z}=\sqrt{m/2 \hbar \omega_{r}}(\omega -\omega _{r})),
\label{eq:IA-dynamic-structure-factor-quasicondensate}
\end{equation}
where $\omega_{r}$ is the recoil frequency and
\begin{equation}
n_{qc}(k_{z})\equiv\int _{-\infty }^{\infty } dz e^{ik_{z}z} \int _{-\infty }^{\infty } dz^{\prime } e^{-ik_{z}z^{\prime }} \sqrt{n_{0}(z)n_{0}(z^{\prime })} e^{-\frac{1}{2}F_{s}(z,z^{\prime })}.
\label{eq:axial-nqc-defined}
\end{equation}

\begin{figure*}[p]
\includegraphics{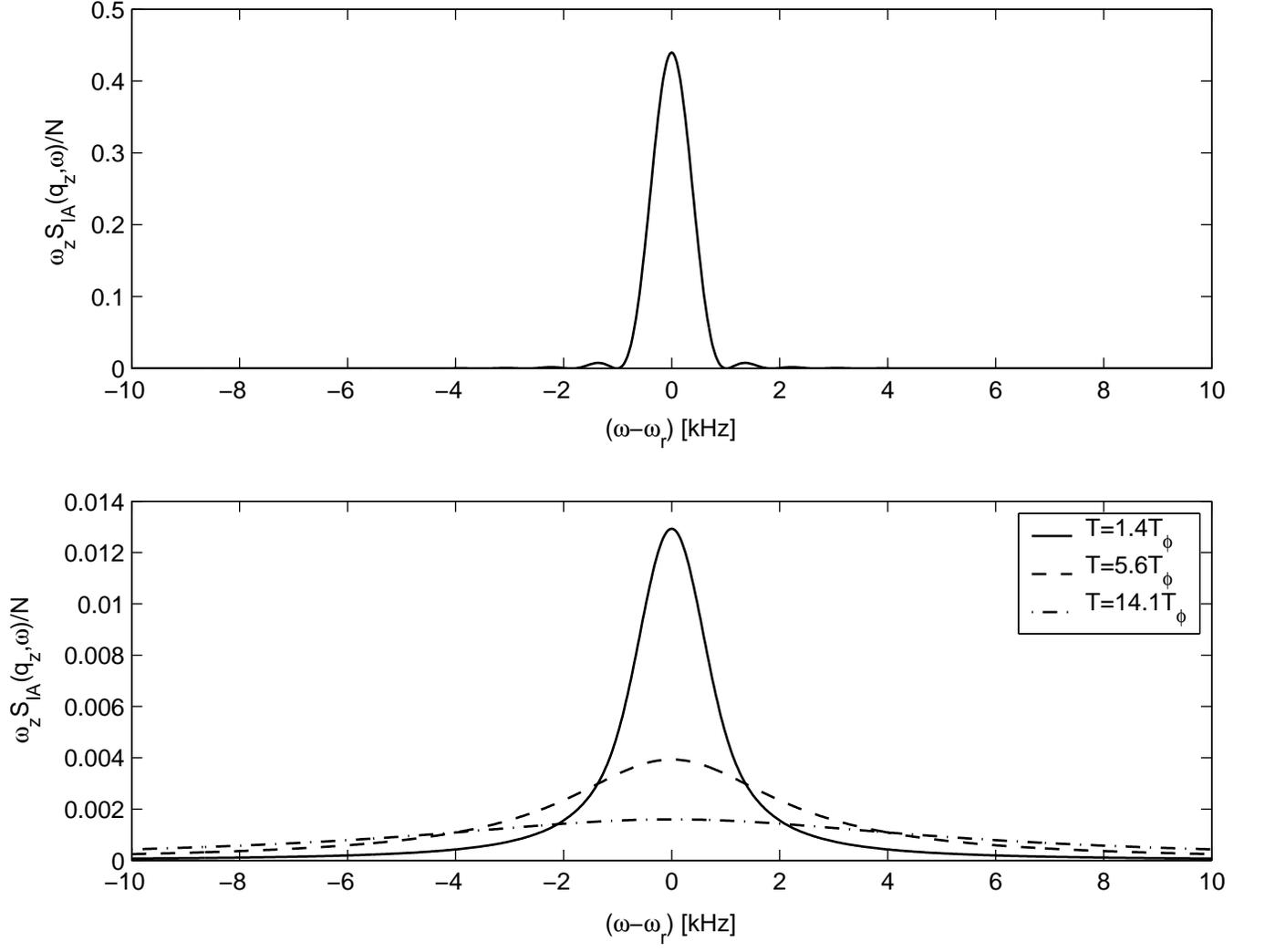}
\caption{The frequency dependence, relative to $\omega_{r}=\hbar q_{z}^{2}/2m$, of the dynamic structure factor in the impulse approximation for a Bose condensate
(top panel) and a quasi-condensate (bottom panel). These results are for $N=10^{4}$, $g_{1D}/\hbar \omega _{z}l _{z}=10$, $k_{B}T_{\phi}/\hbar \omega_{z}=7.1$, $\omega_{z}=2\pi \times 5 \, \mathrm{Hz}$,
$\omega_{\rho}=2\pi \times 10^{4}\, \mathrm{Hz}$ and $\omega _{r}/\omega _{z}=10^{5}$.}
\label{cap:structure-factor}
\end{figure*}

In Fig.~\ref{cap:structure-factor}, we compare $S_{IA}$
for a true Bose condensate and a quasi-condensate [see Eq.~(\ref{eq:IA-dynamic-structure-factor-quasicondensate})] as a function of
the frequency $\omega$, measured relative to the recoil frequency $\omega_{r}$. The width of $S_{IA}$ for the quasi-condensate 
is significantly wider than that of a true Bose condensate, as expected. The resulting axial momentum distribution has Doppler width $\Delta _{IA}$ proportional to $1/l_{\phi }$, which increases (since $l_{\phi }$ decreases) with increasing temperature. For a true condensate,
the Doppler width may be evaluated analytically. Using Eq.~(\ref{eq:IA-dynamic-structure-factor-condensate})
for the IA dynamic structure factor, the Doppler width using Eq.~(\ref{eq:IA-Doppler-width}) is
\begin{equation}
\Delta_{IA}^{c}=\sqrt{2}\frac{\hbar q_{z}}{mZ_{0}}.
\label{eq:impulse-approximation-Doppler-width}
\end{equation}
This Doppler width is inversely proportional to the condensate Thomas-Fermi
axial length $Z_{0}$, which for a true Bose condensate
is the coherence length. Aside from the numerical factor
$\sqrt{2}$, the width is the same as given in Ref.~\cite{zambelli-pra-61-063608} for a condensate
in an anisotropic 3D trap. The fact that the Doppler width is inversely
proportional to the condensate length $Z_{0}$ is intuitively expected.
The Doppler width in Eq.~(\ref{eq:impulse-approximation-Doppler-width})
is a measure of the spread of the momentum distribution of the condensate,
which is inversely proportional to the spatial size of the condensate $Z_{0}$.

\begin{figure*}[p]
\includegraphics{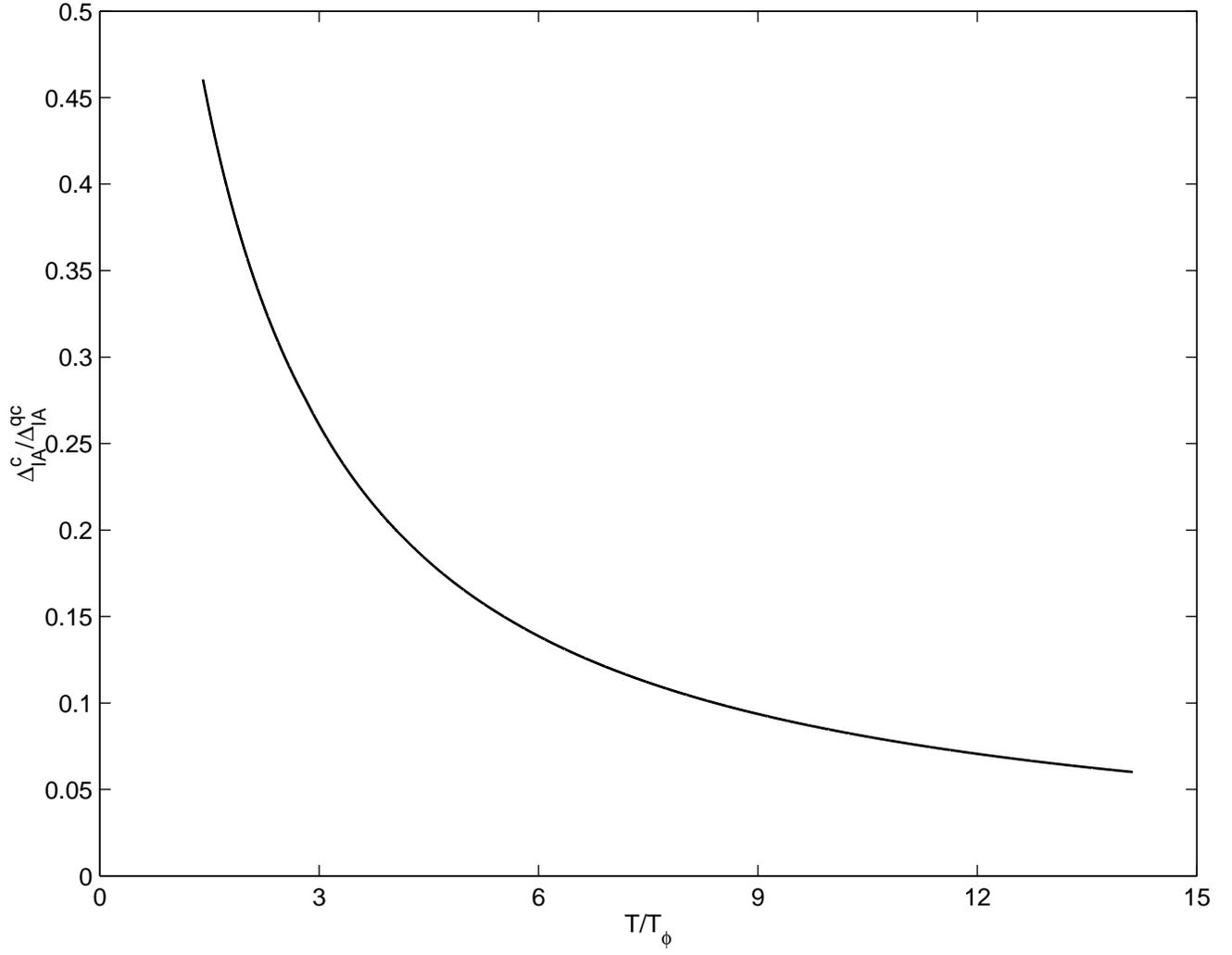}
\caption{Behavior of the correlation length of a quasi-condensate
(Doppler width of the dynamic structure factor) as a function of temperature.
The parameters used are the same as in Fig.~\ref{cap:structure-factor}.}
\label{cap:widths}
\end{figure*}

We can also calculate the Doppler width of the IA dynamic structure factor
of a quasi-condensate, as given by Eq.~(\ref{eq:IA-dynamic-structure-factor-quasicondensate}).
We numerically fit this quasi-condensate IA dynamic structure factor to
the Gaussian expansion of Eq.~(\ref{eq:IA-dynamic-structure-factor-Gaussian-expansion})
at a series of temperatures to obtain $\Delta_{IA}^{qc}$. The results
are given in Fig.~\ref{cap:widths}, where we plot $(\Delta _{IA}^{c}/\Delta _{IA}^{qc})$
versus $T/T_{\phi}$. Since the quantity $(\Delta _{IA}^{c}/\Delta _{IA}^{qc})$
is proportional to $(q_{z}l_{\phi}/m Z_{0})$,
Fig.~\ref{cap:widths} shows the decrease of the correlation length
$l_{\phi}$ of a quasi-condensate as a function of temperature $T$.

In a very recent paper, Gerbier et.~al.~\cite{gerbier-cond-mat-0211094} have given an alternative evaluation of the static phase fluctuation average $F_{s}(z,z^{\prime})$ using a local density approximation (LDA). This has the important advantage that one can obtain a simple analytic expression for $F_{s}(z,z^{\prime})$ in Eq.~(\ref{eq:thermal-phase-argument--static}). They find that $F_{s}(z,z^{\prime})$ is proportional to $|z-z^{\prime}|/\tilde{l}_{\phi}(Z)$, but now the correlation length $\tilde{l}_{\phi}$ depends on the centre of mass axial position variable $Z \equiv \frac{1}{2}(z+z^{\prime})$,
\begin{equation}
\tilde{l}_{\phi}(Z)=l_{\phi}[1-(Z/Z_{0})^{2}]^{2}
\label{eq:centre-of-mass-correlation-length}
\end{equation}
In Ref.~\cite{gerbier-cond-mat-0211094}, it is shown that this simple LDA calculation of $F_{s}(z,z^{\prime})$ is numerically an excellent approximation to the expression analogous to Eq.~(\ref{eq:thermal-phase-argument--static}) for a highly elongated trap for $T\gtrsim 8T_{\phi}$. This is in the temperature range of interest. The analytic expressions developed by Gerbier et.~al.~\cite{gerbier-cond-mat-0211094} for $F_{s}(z,z^{\prime})$ have been successfully used in analyzing their experimental Bragg scattering data \cite{gerbier-cond-mat-0210206} in terms of the altered momentum distribution of a phase-fluctuating 1D quasi-condensate.
\section{\label{sec:Dynamic-correlation-function}Dynamic single-particle correlation function}

We now want to extend the discussion of the well-known results given in Section \ref{sec:Static-correlation-function} and consider the \emph{dynamic} correlation function defined in Eq.~(\ref{eq:phase-fluctuating-single-particle-correlation-function--mean-field-approximation}).
This is the major new result of the present paper. Using Eq.~(\ref{eq:phase-fluctuation-operator--canonical-transformation-solved}),
the exponential argument in Eq.~(\ref{eq:phase-fluctuating-single-particle-correlation-function--mean-field-approximation})
is
\begin{eqnarray}
F(z,z^{\prime };t) & = & \frac{g_{1D}}{Z_{0}}\sum _{j=1}^{\infty }\frac{j+1/2}{2\hbar \omega_{j}}\nonumber \\
 &  & \times [P_{j}^{2}(z/Z_{0})+P_{j}^{2}(z^{\prime }/Z_{0})\nonumber \\
 &  & -2P_{j}(z/Z_{0})P_{j}(z^{\prime }/Z_{0})\cos (\omega _{j}t)]\nonumber \\
 &  & \times [1+2N(\hbar \omega_{j})].
\label{eq:phase-argument--dynamic}
\end{eqnarray}
We can separate the static and dynamic contributions in a natural manner,
\begin{equation}
F(z,z^{\prime };t)=F_{s}(z,z^{\prime })+F_{d}(z,z^{\prime };t),
\label{eq:phase-argument--decomposition-into-static-and-dynamic}
\end{equation}
where the static contribution is given by Eq.~(\ref{eq:thermal-phase-argument--static--evaluated})
and the dynamic contribution is defined by
\begin{equation}
F_{d}(z,z^{\prime };t)\equiv \frac{2g_{1D}k_{B}T}{Z_{0}(\hbar \omega _{z})^{2}}\sum _{j=1}^{\infty }\frac{2j+1}{j(j+1)}P_{j}(z/Z_{0})P_{j}(z^{\prime }/Z_{0})\left[1-\cos (\omega _{j}t)\right].
\label{eq:phase-argument-dynamic--evaluated}
\end{equation}
The dynamic correlation function is then given by
\begin{equation}
C_{\psi ^{\dagger }\psi }(\mathbf{r},\mathbf{r}^{\prime };t)=C_{\psi ^{\dagger }\psi }(\mathbf{r},\mathbf{r}^{\prime };t=0)e^{-\frac{1}{2}F_{d}(z,z^{\prime };t)},
\label{eq:dynamic-single-particle-correlation-function--static-and-dynamic-pieces}
\end{equation}
which has the associated single-particle spectral density
\begin{equation}
C_{\psi ^{\dagger }\psi }(\mathbf{r},\mathbf{r}^{\prime };\omega )=C_{\psi ^{\dagger }\psi }(\mathbf{r},\mathbf{r}^{\prime };t=0)\int _{-\infty }^{\infty }dte^{i\omega t}e^{-\frac{1}{2}F_{d}(z,z^{\prime };t)}.
\label{eq:dynamic-single-particle-correlation-function--Fourier-transform}
\end{equation}

To evaluate the Fourier transform in Eq.~(\ref{eq:dynamic-single-particle-correlation-function--Fourier-transform}),
we make a {}``one-phonon approximation'', as discussed in Ref.~\cite{weling-prb-28-5296}
for a 2D harmonic crystal, and expand $\exp [-\frac{1}{2}F_{d}(z,z^{\prime };t)]$
to first order in $F_{d}$. This approximation is good at isolating
the low frequency (i.e., the long-time) behavior of the dynamic correlation
function. In Ref.~\cite{weling-prb-28-5296}, this kind of first order expansion was shown to give results essentially identical to the exact result in the long wavelength, low frequency region of interest. To lowest order, this expansion gives
\begin{eqnarray}
C_{\psi ^{\dagger }\psi }(\mathbf{r},\mathbf{r}^{\prime };\omega ) & = & C_{\psi ^{\dagger }\psi }(\mathbf{r},\mathbf{r}^{\prime };t=0)\frac{\pi g_{1D}k_{B}T}{Z_{0}(\hbar \omega _{z})^{2}}\nonumber \\
 &  & \times \sum _{j=1}^{\infty }\frac{2j+1}{j(j+1)}P_{j}(z/Z_{0})P_{j}(z^{\prime }/Z_{0})\left[\delta (\omega +\omega _{j})+\delta (\omega -\omega _{j})\right].
\label{eq:dynamic-single-particle-correlation-function--Fourier-transform--evaluated}
\end{eqnarray}
Since the temperature is assumed much larger than the lowest energy
excitations, the energy spacing between the excitations is much smaller
than the temperature. In this high temperature regime,
we may thus approximate the sum by a quasi-classical integration over
the excitation mode frequencies, which gives
\begin{eqnarray}
C_{\psi ^{\dagger }\psi }(\mathbf{r},\mathbf{r}^{\prime };\omega ) & = & \frac{\pi g_{1D}k_{B}T}{Z_{0}(\hbar \omega _{z})^{2}}C_{\psi ^{\dagger }\psi }(\mathbf{r},\mathbf{r}^{\prime };t=0)\nonumber \\
 &  & \times \int _{1}^{\infty }d\nu \frac{2\nu +1}{\nu (\nu +1)}P_{\nu }(z/Z_{0})P_{\nu }(z^{\prime }/Z_{0})\left[\delta (\omega +\omega (\nu ))+\delta (\omega -\omega (\nu ))\right],
\label{eq:dynamic-single-particle-correlation-function--Fourier-transform--continuous}
\end{eqnarray}
where the mode frequencies $\omega(\nu)=\omega_{z}\sqrt{\nu(\nu + 1)/2}$
now depend on the continuous variable $\nu$.

The integral over $\nu $ in 
Eq.~(\ref{eq:dynamic-single-particle-correlation-function--Fourier-transform--continuous})
may be evaluated using the delta function identity
\begin{equation}
\delta (f(x))=\sum _{i}\frac{1}{\left|f^{\prime }(x_{i})\right|}\delta (x-x_{i}),
\label{eq:delta-function-identity}
\end{equation}
where $f(x_{i})=0$ and $f^{\prime }(x_{i})\neq 0$. Using this in Eq.~(\ref{eq:dynamic-single-particle-correlation-function--Fourier-transform--continuous}),
the single-particle dynamic correlation function has the spectral
density at low frequencies
\begin{equation}
C_{\psi ^{\dagger }\psi }(\mathbf{r},\mathbf{r}^{\prime };\omega )=\frac{2\sqrt{2}\pi g_{1D}k_{B}T}{Z_{0}\hbar ^{2}\omega _{z}^{3}}C_{\psi ^{\dagger }\psi }(\mathbf{r},\mathbf{r}^{\prime };t=0)\frac{P_{\nu (\omega )}(z/Z_{0})P_{\nu (\omega )}(z^{\prime }/Z_{0})}{\sqrt{\nu (\omega )[\nu (\omega )+1]}}\theta (\nu (\omega )-1),
\label{eq:dynamic-single-particle-correlation-function--Fourier-transform--approximately-evaluated}
\end{equation}
where 
\begin{equation}
\nu (\omega )\equiv -\frac{1}{2}+\sqrt{\frac{1}{4}+2\frac{\omega ^{2}}{\omega _{z}^{2}}}.
\label{eq:nu(w)}
\end{equation}

The behavior of a \emph{uniform} 1D Bose gas is quite different. A quasi-condensate 
can only form at $T=0$. For non-zero temperatures, the system
cannot be distinguished from a normal non-Bose-condensed gas. As is well known, the key signature of a quasi-condensate
in the $T=0$ uniform 1D Bose gas is that dynamic correlation functions for
low frequencies and momentum have a power law form (see, e.g., Refs.~\cite{weling-prb-28-5296,haldane-prl-47-1840,yip-prl-87-130401})
\begin{equation}
C_{\psi ^{\dagger }\psi }(k,\omega )\sim \frac{1}{\left|\omega ^{2}-c^{2}k^{2}\right|^{1-1/4K}},
\label{eq:correlation-function-uniform-power-law-singularity}
\end{equation}
where $c\equiv \sqrt{gn_{0}/m}$ is the phonon velocity with $n_{0}$
the uniform superfluid density and $K\equiv \pi \hbar \sqrt{n_{0}/mg_{1D}}$
is the Luttinger parameter. As a function of frequency $\omega $,
for low momentum, the spectral density in Eq.~(\ref{eq:correlation-function-uniform-power-law-singularity})
will exhibit a characteristic power law singularity at $\omega=ck$, in place of a delta function $\delta(\omega - ck)$ characteristic of a 3D gas.

\begin{figure*}[p]
\includegraphics{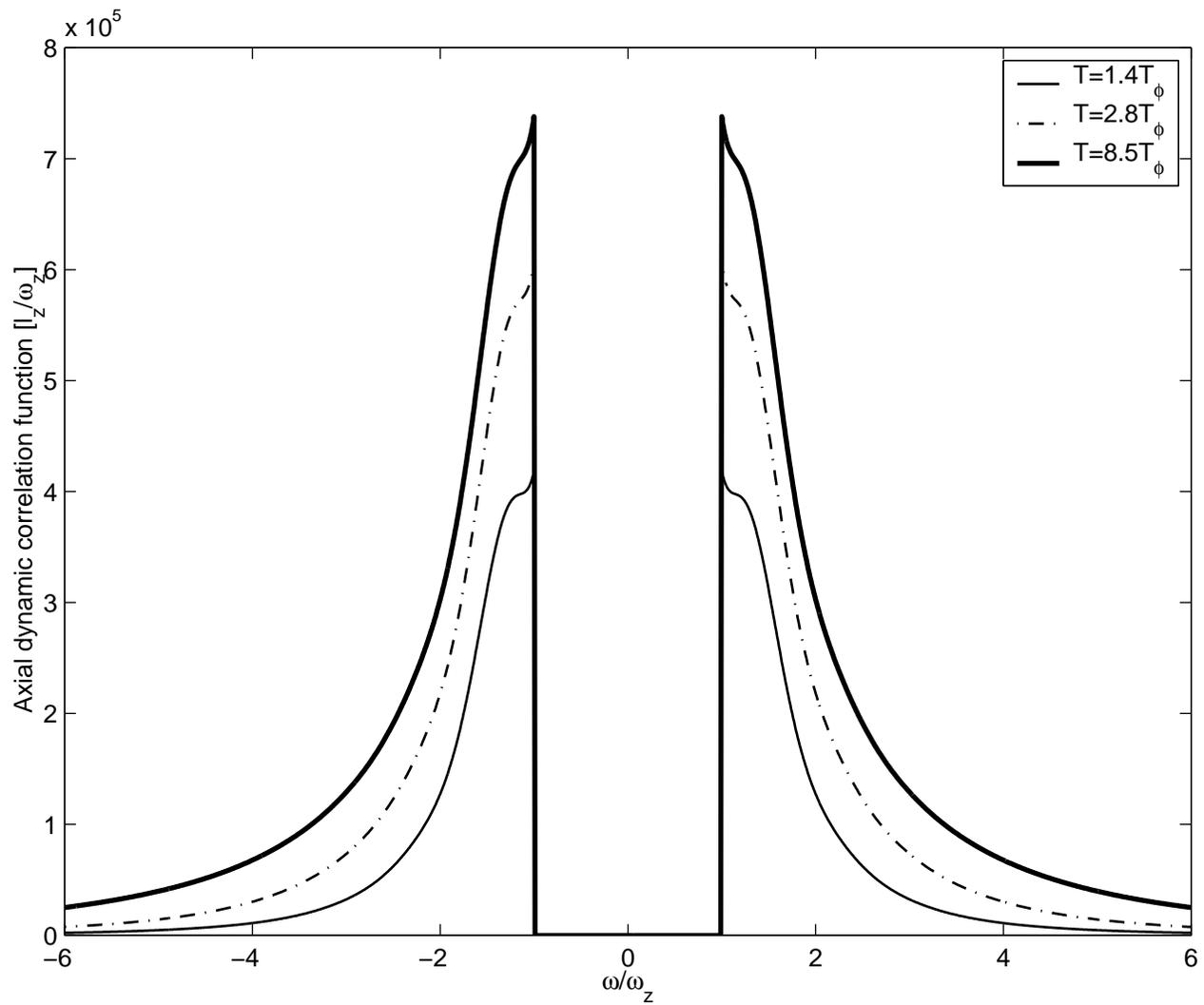}
\caption{Axial dynamic correlation
function for a quasi-condensate [defined in Eq.~(\ref{eq:axial-single-particle-correlation-function-defined-quasi-condensate})].
The parameters $N$, $g_{1D}$, $T_{\phi}$, $\omega_{z}$ and $\omega_{\rho }$
are the same as in Fig.~\ref{cap:structure-factor}, with $k_{z}=0.01/l_{z}$.}
\label{cap:dynamic-correlation-function}
\end{figure*}

The expression for the dynamic correlation function in Eq.~(\ref{eq:dynamic-single-particle-correlation-function--Fourier-transform--approximately-evaluated}) only describes long wavelength excitations. As in the case of the static
correlation function discussed in Section \ref{sec:Static-correlation-function}, we consider the Fourier transformed dynamic correlation function
\begin{equation}
C_{\psi ^{\dagger }\psi }(-\mathbf{k},\mathbf{k};\omega)\equiv \int d\mathbf{r}e^{i\mathbf{k}\cdot \mathbf{r}}\int d\mathbf{r}^{\prime }e^{-i\mathbf{k}\cdot \mathbf{r}^{\prime }}C_{\psi ^{\dagger }\psi }(\mathbf{r},\mathbf{r}^{\prime };\omega ).
\label{eq:dynamic-single-particle-correlation-function--momentum-and-frequency-space}
\end{equation}
Substituting the expression in Eq.~(\ref{eq:dynamic-single-particle-correlation-function--Fourier-transform--approximately-evaluated})
into Eq.~(\ref{eq:dynamic-single-particle-correlation-function--momentum-and-frequency-space}),
gives
\begin{equation}
C_{\psi ^{\dagger }\psi }(-\mathbf{k},\mathbf{k};\omega)=4\pi l_{\rho }^{2}e^{-l_{\rho }^{2}(k_{x}^{2}+k_{y}^{2})}C^{qc}(k_{z};\omega ),
\label{eq:dynamic-single-particle-correlation-function--momentum-and-frequency-space--evaluated}
\end{equation}
where the quasi-condensate axial single-particle correlation function is defined by
\begin{eqnarray}
C^{qc}(k_{z};\omega) & \equiv  & \frac{1}{\omega _{z}}\frac{2\sqrt{2}\pi g_{1D}k_{B}T}{Z_{0}(\hbar \omega _{z})^{2}}\int _{-\infty }^{\infty }dz\int _{-\infty }^{\infty }dz^{\prime }e^{ik_{z}(z-z^{\prime })}\nonumber \\
 &  & \times \sqrt{n_{0}(z)n_{0}(z^{\prime })}e^{-\frac{1}{2}F_{s}(z,z^{\prime })}\frac{P_{\nu (\omega )}(z/Z_{0})P_{\nu (\omega )}(z^{\prime }/Z_{0})}{\sqrt{\nu (\omega )[\nu (\omega )+1]}}\theta (\nu (\omega )-1).
\label{eq:axial-single-particle-correlation-function-defined-quasi-condensate}
\end{eqnarray}
The integrals over the axial directions $z$ and $z^{\prime }$ in
Eq.~(\ref{eq:dynamic-single-particle-correlation-function--momentum-and-frequency-space--evaluated})
have been evaluated numerically, with the results shown in
Fig.~\ref{cap:dynamic-correlation-function}.

For comparison, we also calculate the correlation function corresponding
to an ordinary Bose condensate at $T=0$. As in our calculations of
the dynamic correlation function for a quasi-condensate, we ignore
density fluctuations. The correlation function for a Bose condensate
in a 1D trap is given by {[}see Eqs.~(\ref{eq:radial-wavefunction})
and (\ref{eq:axial-condensate-density}){]}
\begin{equation}
C_{\psi ^{\dagger }\psi }(\mathbf{r},\mathbf{r};t)=\phi _{\rho }(x,y)\phi _{\rho }(x^{\prime },y^{\prime })\sqrt{n_{0}(z)n_{0}(z^{\prime })},
\label{eq:dynamic-single-particle-correlation-function--BEC}
\end{equation}
so that the corresponding momentum and frequency dependent dynamic
correlation function is
\begin{equation}
C_{\psi ^{\dagger }\psi }(\mathbf{k},-\mathbf{k};\omega )=4 \pi l_{\rho }^{2}e^{-l_{\rho }^{2}(k_{x}^{2}+k_{y}^{2})}C^{c}(k_{z};\omega),
\label{eq:dynamic-single-particle-correlation-function--BEC--momentum-and-frequency-space}
\end{equation}
where the condensate axial single-particle correlation function is defined by
\begin{equation}
C^{c}(k_{z};\omega)\equiv \frac{2\pi^{3}\mu_{0}}{g_{1D}}\frac{J_{1}^{2}(Z_{0}k_{z})}{k_{z}^{2}}\delta(\omega).
\label{eq:axial-single-particle-correlation-function-defined-BEC}
\end{equation}
This pure condensate result is strongly peaked about zero frequency,
i.e., a delta function $\delta(\omega)$.

Because of the step function in Eq.~(\ref{eq:dynamic-single-particle-correlation-function--Fourier-transform--approximately-evaluated}),
the dynamic correlation function will have no contribution from frequencies
for which $\nu (\omega )<1$, which corresponds to frequencies
$\left|\omega \right|$ less than the axial trap frequency $\omega_{z}$. This gap clearly arises because of the finite-size of the system. All frequencies below the
lowest excitation energy have zero-weight in the dynamic correlation
function, as expected. The same gap would, of course, also occur in the single-particle spectral density of a non-Bose condensed 1D gas \emph{above} $T_{BEC}$. In contrast,
in the case of a trapped Bose-condensed gas, there is still a large contribution
at zero frequency arising from the condensate
itself. Thus, a trapped 1D quasi-condensate exhibits the absence
of true long-range order through the presence of a gap in
the correlation function spectrum for frequencies below that of the
lowest energy collective mode ($\omega _{z}$), just as a non-Bose condensed gas would. This effect also occurs
for highly elongated traps, whose static correlation properties were
considered in Ref.~\cite{petrov-prl-87-050404}. As we noted earlier, aside
from minor numerical differences arising from different wavefunctions for
the low energy phase fluctuations, the behavior of a highly elongated trap is
the same as the 1D system we are considering in this paper.

In the present section, we have calculated the dynamic 1D single-particle correlation function directly from the expression in Eq.~(\ref{eq:dynamic-single-particle-correlation-function--Fourier-transform--evaluated}), approximating the discrete sum over the excitations by an integration. This does not give any simple analytic expression analogous to Eq.~(\ref{eq:correlation-function-uniform-power-law-singularity}) which clearly exhibits the properties of a quasi-condensate. An alternative might be to use the LDA approach which Gerbier et.~al.~\cite{gerbier-cond-mat-0211094} have recently used to evaluate the static correlation functions in a highly elongated trapped Bose gas in the high temperature limit (see discussion at the end of Section \ref{sec:Static-correlation-function}).
\section{\label{sec:Out-coupling-experiment}Out-coupling experiment}

It has been recently shown \cite{luxat-pra-65-043618} that an out-coupling
experiment, based upon two-photon Raman scattering, provides a way
of measuring the single-particle correlation functions of a Bose-condensed
gas. In this section, we apply the formalism of Ref.~\cite{luxat-pra-65-043618}
(see also Refs.~\cite{japha-prl-82-1079,choi-pra-61-063606}) to
discuss how a Raman scattering experiment might be used to probe the
\emph{dynamic} single-particle correlation function of a 1D quasi-condensate.
So far, experimental interest has focussed upon the \emph{static}
correlation function discussed in Section III. We discuss an experiment involving out-coupling of atoms from a trapped hyperfine state to an untrapped hyperfine state.

As in Ref.~\cite{luxat-pra-65-043618}, two external laser fields
produce a Raman transition from a trapped hyperfine to a second hyperfine
state. Atoms in the trapped hypefine state are labelled by {}``1''
and those in the untrapped atomic hyperfine state are labelled by
{}``2''. The energies of the trapped and untrapped hyperfine states
are denoted by $\varepsilon _{1}$ and $\varepsilon _{2}$, respectively.
The difference in energy between these two hyperfine states is denoted
by $\Delta \varepsilon \equiv \varepsilon _{2}-\varepsilon _{1}$. The chemical potentials
of the trapped and untrapped gases are denoted as $\mu _{1}$ and $\mu _{2}$, respectively. The difference in the chemical potentials is denoted as $\Delta \mu \equiv \mu _{2}-\mu _{1}$.
Similarly, the annihilation operator $\hat{\psi }_{1(2)}$ destroys
an atom in the hyperfine state 1(2). The external potential that describes
the perturbation of the laser fields can then be written as
\begin{equation}
\hat{V}_{ext}(t)=\gamma \int d\mathbf{r}\left[e^{i\mathbf{q}\cdot \mathbf{r}-i\omega t}\hat{\psi }_{1}(\mathbf{r})\hat{\psi }_{2}^{\dagger }(\mathbf{r})+H.c.\right],
\label{eq:external-potential}
\end{equation}
where $\gamma$ denotes the laser field strength, while $\mathbf{q}\equiv \mathbf{k}_{a}-\mathbf{k}_{b}$ is the wave vector difference and $\omega \equiv \omega _{a}-\omega _{b}$ is the frequency difference between the two laser beams. We set $\hbar=1$ in this section. 

In Ref.~\cite{luxat-pra-65-043618}, it is shown that the current
of atoms flowing from a trapped gas to an untrapped gas induced
by the external perturbation in Eq.~(\ref{eq:external-potential})
is given by (within linear response theory)
\begin{eqnarray}
\delta I_{out}(\mathbf{q},\omega ) & = & 2\gamma ^{2}\mathrm{Re}\int d\mathbf{r}\int d\mathbf{r}^{\prime }\int _{-\infty }^{\infty }dte^{i\mathbf{q}\cdot (\mathbf{r}-\mathbf{r}^{\prime })}e^{-\eta t}\theta (t)e^{-i(\omega -\Delta \varepsilon -\Delta \mu)t}\nonumber \\
 &  & \times \left[C_{\psi _{1}^{\dagger }\psi _{1}}(\mathbf{r}^{\prime },\mathbf{r};-t)C_{\psi _{2}\psi _{2}^{\dagger }}(\mathbf{r}^{\prime },\mathbf{r};-t)-C_{\psi _{1}\psi _{1}^{\dagger }}(\mathbf{r},\mathbf{r}^{\prime };t)C_{\psi _{2}^{\dagger }\psi _{2}}(\mathbf{r},\mathbf{r}^{\prime };t)\right].
\label{eq:out-coupling-current-in-time}
\end{eqnarray}
We use the notation
\begin{equation}
C_{AB}(\mathbf{r},\mathbf{r}^{\prime };t)\equiv \left\langle \hat{A}(\mathbf{r},t)\hat{B}(\mathbf{r}^{\prime })\right\rangle \label{eq:correlation-function-definition}
\end{equation}
to describe the correlation function of two operators $\hat{A}$ and
$\hat{B}$ in the interaction picture, with the grand-canonical Hamiltonian
$\hat{K}\equiv \hat{H}-\mu _{1}\hat{N}_{1}-\mu _{2}\hat{N}_{2}$ determining
the time-evolution of operators. The operator $\hat{N}_{1(2)}$
is the number operator for atoms in hyperfine state labelled by 1(2).

We may rewrite Eq.~(\ref{eq:out-coupling-current-in-time}) for the
out-coupling current in terms of Fourier transformed single-particle
correlation functions \cite{luxat-pra-65-043618}
\begin{equation}
\delta I_{out}(\mathbf{q},\omega )=\delta I_{out}^{(a)}(\mathbf{q},\omega )+\delta I_{out}^{(b)}(\mathbf{q},\omega ),
\label{eq:out-coupling-current-split-into-pieces}
\end{equation}
where the two contributions $\delta I_{out}^{(a)}$ and $\delta I_{out}^{(b)}$
to the out-coupling atom current are given by
\begin{eqnarray}
\delta I_{out}^{(a)}(\mathbf{q},\omega ) & = & \gamma ^{2}\mathrm{Re}\int d\mathbf{r}\int d\mathbf{r}^{\prime }e^{i\mathbf{q}\cdot (\mathbf{r}-\mathbf{r}^{\prime })}\int _{-\infty }^{\infty }\frac{d\omega ^{\prime }}{2\pi }\nonumber \\
 &  & \times [C_{\psi _{1}^{\dagger }\psi _{1}}(\mathbf{r}^{\prime },\mathbf{r};-\omega ^{\prime })C_{\psi _{2}\psi _{2}^{\dagger }}(\mathbf{r}^{\prime },\mathbf{r};\omega ^{\prime }+\omega -\Delta \varepsilon -\Delta \mu)\nonumber \\
 &  & \qquad -C_{\psi _{1}\psi _{1}^{\dagger }}(\mathbf{r},\mathbf{r}^{\prime };\omega ^{\prime })C_{\psi _{2}^{\dagger }\psi _{2}}(\mathbf{r},\mathbf{r}^{\prime };-\omega ^{\prime }-\omega +\Delta \varepsilon + \Delta \mu)]
\label{eq:out-coupling-current-a}
\end{eqnarray}
and
\begin{eqnarray}
\delta I_{out}^{(b)}(\mathbf{q},\omega ) & = & -2\gamma ^{2}\mathrm{Im}\int d\mathbf{r}\int d\mathbf{r}^{\prime }e^{i\mathbf{q}\cdot (\mathbf{r}-\mathbf{r}^{\prime })}\mathrm{P}\int _{-\infty }^{\infty }\frac{d\omega ^{\prime }}{2\pi }\int _{-\infty }^{\infty }\frac{d\omega ^{\prime \prime }}{2\pi }\nonumber \\
 &  & \times \frac{C_{\psi _{1}^{\dagger }\psi _{1}}(\mathbf{r}^{\prime },\mathbf{r};-\omega ^{\prime })C_{\psi _{2}\psi _{2}^{\dagger }}(\mathbf{r}^{\prime },\mathbf{r};\omega ^{\prime \prime })-C_{\psi _{1}\psi _{1}^{\dagger }}(\mathbf{r},\mathbf{r}^{\prime };\omega ^{\prime })C_{\psi _{2}^{\dagger }\psi _{2}}(\mathbf{r},\mathbf{r}^{\prime };-\omega ^{\prime \prime })}{\omega ^{\prime \prime }-\omega ^{\prime }-\omega +\Delta \varepsilon + \Delta \mu },
\label{eq:out-coupling-current-b}
\end{eqnarray}
where the symbol $\mathrm{P}$ refers to principal value integration. The single-particle correlation functions are even in time and thus
their Fourier transform is consequently real. The only component of
the integrand in Eq.~(\ref{eq:out-coupling-current-b}) for $\delta I_{out}^{(b)}$
which is imaginary is the term proportional to $\sin [\mathbf{q}\cdot (\mathbf{r}-\mathbf{r}^{\prime })]$. Since the trapping potential is invariant under spatial inversions $\mathbf{r}\rightarrow -\mathbf{r}$, the correlation functions are
also invariant under spatial inversions. The integrals over $\mathbf{r}$
and $\mathbf{r}^{\prime }$ in Eq.~(\ref{eq:out-coupling-current-b})
thus vanish in the imaginary term, since $\sin [\mathbf{q}\cdot (\mathbf{r}-\mathbf{r}^{\prime })]$ is odd under spatial inversion. As a result the contribution $\delta I_{out}^{(b)}$
is zero, and we are left with $\delta I_{out}^{(a)}$ in Eq.~(\ref{eq:out-coupling-current-a}).

We may rewrite the expression for the out-coupling current in Eq.~(\ref{eq:out-coupling-current-a}) in terms of the
spectral density of the second gas $A_{\psi _{2}^{\dagger }\psi _{2}}(\mathbf{r},\mathbf{r}^{\prime };\omega )$
using the well-known identity
\begin{equation}
C_{\psi _{2}^{\dagger }\psi _{2}}(\mathbf{r},\mathbf{r}^{\prime };-\omega )=N_{2}(\omega)A_{\psi _{2}^{\dagger }\psi _{2}}(\mathbf{r},\mathbf{r}^{\prime };\omega ),
\label{eq:correlation-function-in-terms-of-spectral-density}
\end{equation}
where $N_{2}$ is the Bose distribution function for the untrapped
gas evaluated at temperature $T_{2}$. For a dilute, classical gas,
one has $N_{2}(\omega)\ll 1$ and the expression in Eq.~(\ref{eq:out-coupling-current-a})
reduces to
\begin{equation}
\delta I_{out}(\mathbf{q},\omega )=\gamma ^{2}\mathrm{Re}\int d\mathbf{r}\int d\mathbf{r}^{\prime }e^{i\mathbf{q}\cdot (\mathbf{r}-\mathbf{r}^{\prime })}\int _{-\infty }^{\infty }\frac{d\omega ^{\prime }}{2\pi }C_{\psi _{1}^{\dagger }\psi _{1}}(\mathbf{r}^{\prime },\mathbf{r};-\omega ^{\prime })A_{\psi _{2}^{\dagger }\psi _{2}}(\mathbf{r},\mathbf{r}^{\prime };\omega ^{\prime }+\omega -\Delta \varepsilon - \Delta \mu).
\label{eq:out-coupling-current-in-terms-of-spectral-density}
\end{equation}
The atoms in the second hyperfine state can be described as a uniform, non-interacting 3D gas \cite{luxat-pra-65-043618}. Using the uniform gas spectral density
\begin{equation}
A_{\psi _{2}^{\dagger }\psi _{2}}(\mathbf{k},\omega )=2\pi \delta (\omega -\tilde{\varepsilon }_{\mathbf{k}2}),
\label{eq:spectral-density-for-uniform,ideal-gas}
\end{equation}
where $\tilde{\varepsilon }_{\mathbf{k}2}\equiv k^{2}/2m-\mu _{2}=\varepsilon_{\mathbf{k}}-\mu _{2}$
is the excitation energy of an out-coupled, untrapped atom, the linear response current $\delta I_{out}$ reduces to
\begin{equation}
\delta I_{out}(\mathbf{q},\tilde{\omega} )=\gamma ^{2}\mathrm{Re}\int \frac{d\mathbf{k}}{(2\pi )^{3}}C_{\psi ^{\dagger }\psi }(-(\mathbf{k}+\mathbf{q}),(\mathbf{k}+\mathbf{q});\tilde{\omega} +\mu _{0} - \varepsilon _{\mathbf{k}}).
\label{eq:out-coupling-current-in-terms-of-only-single-particle-correlation-function}
\end{equation}
We have introduced the {}``detuning parameter'' $\tilde{\omega} \equiv \omega -\Delta \varepsilon$ and for notational convenience, we have also dropped the distinction between
trapped and untrapped hyperfine states in Eq.~(\ref{eq:out-coupling-current-in-terms-of-only-single-particle-correlation-function}).
The dynamic single-particle correlation function and chemical potential
$\mu _{0}$ [see Eq.~(\ref{eq:condensate-chemical-potential})] in Eq.~(\ref{eq:out-coupling-current-in-terms-of-only-single-particle-correlation-function})
refer to the trapped gas. The result in Eq.~(\ref{eq:out-coupling-current-in-terms-of-only-single-particle-correlation-function})
shows how the out-coupling atomic current is formally given directly in terms of the dynamic single-particle correlation function of the trapped gas \cite{luxat-pra-65-043618}. The expression in Eq.~(\ref{eq:out-coupling-current-in-terms-of-only-single-particle-correlation-function}) applies equally to a true Bose condensate and a quasi-condensate.

We first calculate the out-coupling current arising from a 1D Bose condensate at $T=0$. We neglect the contribution from the noncondensate atoms (the thermal cloud), as in our discussion of the quasi-condensate in Section \ref{sec:Dynamic-correlation-function}. Using
Eq.~(\ref{eq:dynamic-single-particle-correlation-function--BEC--momentum-and-frequency-space})
for the dynamic correlation function of a Bose condensate, the
out-coupling current in Eq.~(\ref{eq:out-coupling-current-in-terms-of-only-single-particle-correlation-function}) is
\begin{equation}
\delta I_{out}^{c}(\mathbf{q},\tilde{\omega} )=4 \pi l_{\rho}^{2} \gamma^{2} \mathrm{Re} \int \frac{d\mathbf{k}}{(2\pi)^{3}}e^{-l_{\rho }^{2}[(k_{x}+q_{x})^{2}+(k_{y}+q_{y})^{2}]}C^{c}(k_{z}+q_{z};\tilde{\omega}+\mu_{0} - \varepsilon_{\mathbf{k}}),
\label{eq:out-coupling-current-for-BEC}
\end{equation}
where $C^{c}$ is the condensate axial single-particle correlation function defined in Eq.~(\ref{eq:axial-single-particle-correlation-function-defined-BEC}). Similarly, using Eq.~(\ref{eq:dynamic-single-particle-correlation-function--momentum-and-frequency-space--evaluated}) for the dynamic correlation function of a quasi-condensate, we find that the out-coupling current from a quasi-condensate is
\begin{equation}
\delta I_{out}^{qc}(\mathbf{q},\tilde{\omega})=4\pi l_{\rho}^{2} \gamma^{2} \mathrm{Re}\int \frac{d\mathbf{k}}{(2\pi)^{3}}e^{-l_{\rho}^{2}\left[(k_{x}+q_{x})^{2}+( k_{y}+q_{y})^{2}\right] }C^{qc}(k_{z}+q_{z};\tilde{\omega}+\mu_{0} - \varepsilon_{\mathbf{k}}),
\label{eq:out-coupling-current-for-quasi-condensate}
\end{equation}
where $C^{qc}$ is the quasi-condensate axial single-particle correlation function defined in Eq.~(\ref{eq:axial-single-particle-correlation-function-defined-quasi-condensate}).

Since the radial confinement of the trapped gas is significant, the
momentum distribution in the radial direction is very broad. The low
frequency, long wavelength fluctuations in the axial direction are
difficult to resolve in the total out-coupling current. It is better to measure the out-coupling current from atoms with low momentum. Our results in Section \ref{sec:Dynamic-correlation-function} for the dynamic single-particle correlation functions in Eq.~(\ref{eq:dynamic-single-particle-correlation-function--BEC--momentum-and-frequency-space}) and Eq.~(\ref{eq:dynamic-single-particle-correlation-function--momentum-and-frequency-space--evaluated})
can be used to calculate the momentum-resolved atom current
for a quasi-condensate, and then compared to that for a true Bose condensate. The momentum-resolved out-coupling current from a trapped Bose condensate, using Eq.~(\ref{eq:out-coupling-current-for-BEC}), is
\begin{equation}
\delta I_{out}^{c}(\mathbf{q},\mathbf{k};\tilde{\omega})=\frac{8\pi ^{4}l_{\rho }^{2}\mu _{0}\gamma ^{2}}{g_{1D}} e^{-l_{\rho }^{2}(k_{x}^{2}+k_{y}^{2})}\frac{J_{1}^{2}(Z_{0}k_{z})}{k_{z}^{2}}\delta (\tilde{\omega} +\mu _{0} - \varepsilon _{\mathbf{k}+\mathbf{q}}),
\label{eq:momentum-resolved-out-coupling-current-for-BEC}
\end{equation}
which is strongly peaked (i.e., a delta function) about $\tilde{\omega}\equiv \omega-\Delta\varepsilon=\varepsilon_{\mathbf{k}+\mathbf{q}}-\mu_{0}$. In contrast, from Eq.~(\ref{eq:out-coupling-current-for-quasi-condensate}), the out-coupling current from a trapped quasi-condensate is
\begin{equation}
\delta I_{out}^{qc}(\mathbf{q},\mathbf{k};\tilde{\omega})=4\pi \gamma^{2} l_{\rho}^{2}\mathrm{Re}\,e^{-l_{\rho}^{2}(k_{x}^{2}+k_{y}^{2}) }C^{qc}(k_{z};\tilde{\omega}+\mu_{0}-\varepsilon_{\mathbf{k}+\mathbf{q}}).
\label{eq:momentum-resolved-out-coupling-current-for-quasi-condensate}
\end{equation}
Using Eq.~(\ref{eq:axial-single-particle-correlation-function-defined-quasi-condensate}) for $C^{qc}$, we observe that there will be no contribution to the momentum-resolved out-coupling current for $\left|\tilde{\omega}+\mu_{0}-\varepsilon_{\mathbf{k}+\mathbf{q}} \right|<\omega_{z}$, in contrast to the atom current in Eq.~(\ref{eq:momentum-resolved-out-coupling-current-for-BEC}) from a Bose condensate.

\begin{figure*}[p]
\includegraphics{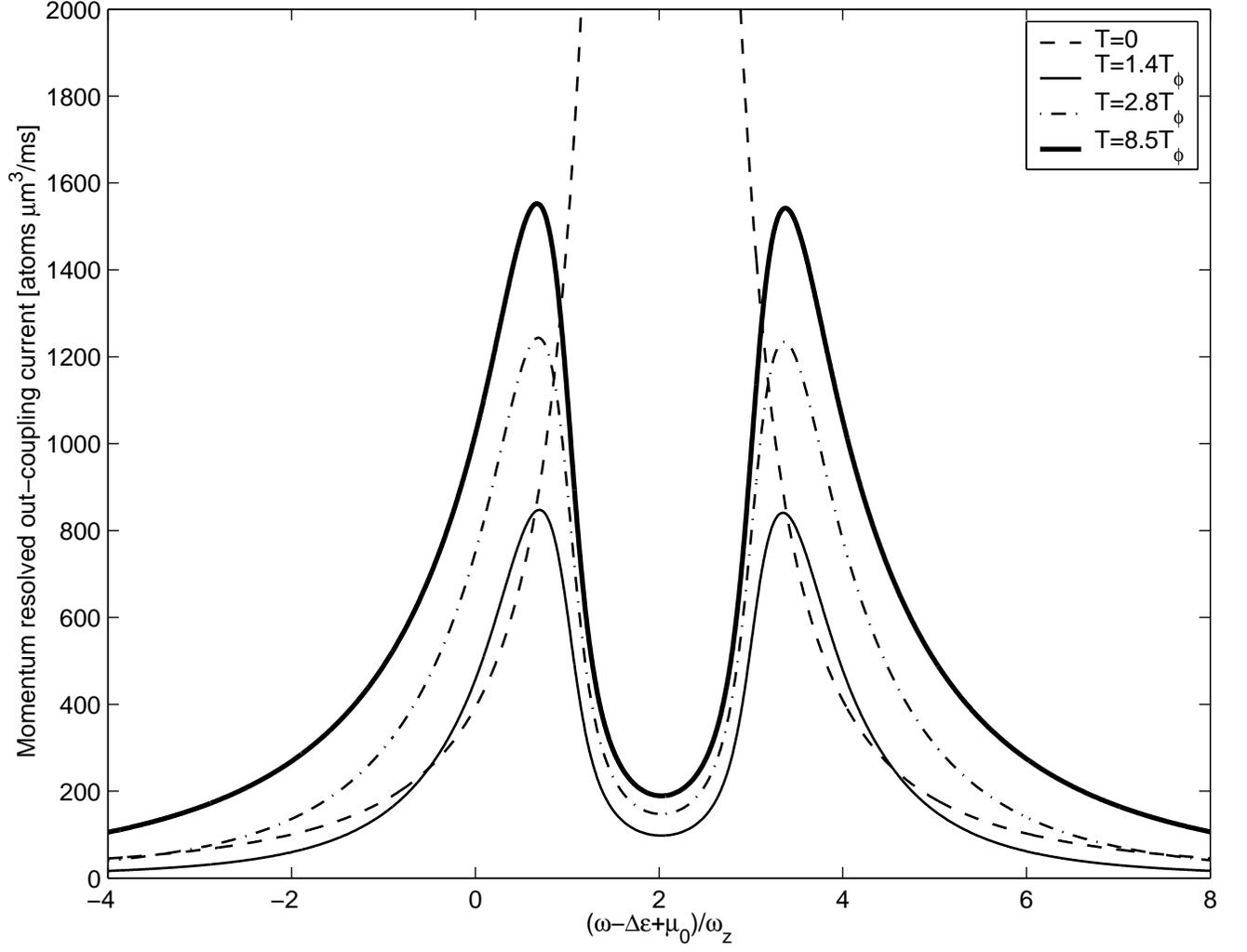}
\caption{Momentum resolved
out-coupling current. The parameters $N$, $g_{1D}$, $T_{\phi}$, $\omega _{z}$ and $\omega _{\rho }$ are the same as in Fig.~\ref{cap:structure-factor},
with $k_{x}=k_{y}=k_{z}=0.01/l_{z}$, $q_{x}=2/l_{z}$, $q_{y}=q_{z}=0$ and $\gamma=0.4 \omega_{z}$. The true $T=0$ condensate current is the dashed line. The results have been Lorentzian broadened by $\Gamma/\omega_{z}=0.25$ to reflect finite energy instrumental resolution.}
\label{cap:momentum-resolved}
\end{figure*}

Fig.~\ref{cap:momentum-resolved} shows our results for the momentum-resolved out-coupling current from a true Bose condensate and a quasi-condensate, both broadened to take into account energy resolution. There is a gap in the quasi-condensate momentum-resolved current centred about $\varepsilon_{\mathbf{k}+\mathbf{q}}$, as there would be in the case of a quasi-1D Bose gas in a highly elongated cigar trap (see the discussion in Section \ref{sec:Dynamic-correlation-function}). If a coherent Bose condensate is present, then there is a significant out-coupling current in the centre of this gap arising from the tunneling of atoms from the condensate (see Fig.~\ref{cap:momentum-resolved}). The gap in the quasi-condensate momentum-resolved out-coupling current arises simply because the excitations have a minimum value given by the axial trap frequency $\omega_{z}$.
\section{\label{sec:Conclusions}Conclusions}

In this paper we have studied both the static and dynamic single-particle
correlations in 1D trapped Bose gases. As is well known from studies
in uniform systems, the unique physics of the 1D Bose gas emerges
because of the long wavelength, low energy fluctuations of the phase,
which destroy the long-range order of the Bose condensate. This leads
to distinctive features in the single-particle correlation functions. In a
uniform Bose gas, the key signature of a quasi-condensate is that single-particle
correlations exhibit power law behavior at low frequencies and long wavelengths. 

Recently there has been considerable attention paid to how the \emph{static} single-particle correlation functions are modified in 1D trapped Bose gases \cite{petrov-prl-84-2551,petrov-prl-85-3745,petrov-prl-87-050404,shlyapnikov-cargese-2-407,gerbier-cond-mat-0211094} as a result of phase fluctuations. These predictions have been largely confirmed by recent experiments \cite{dettmer-prl-87-160406,gerbier-cond-mat-0210206}. In the present paper, we have reviewed the calculation of the single-particle correlation function in 1D quasi-condensates, with emphasis on the momentum distribution of atoms and how this can be measured using Doppler-sensitive Bragg scattering experiments \cite{gerbier-cond-mat-0210206}. Our discussion of the \emph{static} single-particle correlation functions in Section \ref{sec:Static-correlation-function} is needed as background for the generalization in Section \ref{sec:Dynamic-correlation-function} to \emph{time-dependent} single-particle correlation functions. We also show that both the static and dynamic correlation functions can be evaluated using the excitations given by a quantized 1D hydrodynamic model \cite{wu-pra-54-4204}, as a simpler alternative to previous discussions.

In Section \ref{sec:Out-coupling-experiment}, we review a recent theory of out-coupled atom currents tunneling from trapped Bose gases \cite{japha-prl-82-1079,choi-pra-61-063606,luxat-pra-65-043618}. This kind of two-photon Raman scattering experiment might be a way of measuring the frequency dependent single-particle spectral density of a 1D quasi-condensate. This would be a natural extension of the use of Bragg scattering to study the static momentum distribution of quasi-condensates \cite{gerbier-cond-mat-0210206}.

\appendix*
\section{}

The impulse approximation for the dynamic structure factor $S(\mathbf{q},\omega)$ is often written down without much discussion. The physics behind the IA used in reducing Eq.~(\ref{eq:dynamic-structure-factor-momentum-representation}) to Eq.~(\ref{eq:IA-dynamic-structure-factor}) is that for very large momentum and energy transfers, the motion of the atoms from their initial equilibrium distribution is only probed over very small distances and times. Thus the atom dynamics can be treated as if the atoms were free particles. Taking ($\hbar=1$)
\begin{equation}
\hat{\psi}(\mathbf{k}+\mathbf{q},t)=e^{-i E_{\mathbf{k}+\mathbf{q}} t} \hat{\psi}(\mathbf{k}+\mathbf{q},0)
\label{eq:appendix-ckq}
\end{equation}
\begin{equation}
\hat{\psi}^{\dagger}(\mathbf{k},t)=e^{i E_{\mathbf{k}} t} \hat{\psi}^{\dagger}(\mathbf{k},0),
\label{eq:appendix-ck}
\end{equation}
one sees that Eq.~(\ref{eq:dynamic-structure-factor-momentum-representation}) reduces to
\begin{equation}
S(\mathbf{q},\omega)=\int \frac{d\mathbf{k} d\mathbf{k}^{\prime}}{(2\pi)^{3}} \delta \left( \omega - (E_{\mathbf{k}+\mathbf{q}} - E_{\mathbf{k}}) \right) \left\langle \hat{\psi}^{\dagger}(\mathbf{k}) \hat{\psi}(\mathbf{k}+\mathbf{q}) \hat{\psi}^{\dagger}(\mathbf{k}^{\prime}) \hat{\psi}(\mathbf{k}^{\prime}-\mathbf{q}) \right\rangle.
\label{eq:appendix-dynamic-structure-factor}
\end{equation}
In the IA, since $q$ is very large, one can approximate the final energy $E_{\mathbf{k}+\mathbf{q}}$ by the free atom kinetic energy $\varepsilon_{\mathbf{k}+\mathbf{q}} \equiv (\mathbf{k}+\mathbf{q})^{2}/2m$. We do not necessarily know the initial energy $E_{\mathbf{k}}$ (which may be complex) of a typical atom in the equilibrium distribution. However, clearly one has  $E_{\mathbf{k}+\mathbf{q}} \gg E_{\mathbf{k}}$ and thus $E_{\mathbf{k}}$ is negligible. Moreover, the energy conservation factor in Eq.~(\ref{eq:appendix-dynamic-structure-factor}) can be reduced to
\begin{equation}
\delta (\omega - q^{2}/2m - \mathbf{q} \cdot \mathbf{k}/m),
\label{eq:appendix-delta-function-reduced}
\end{equation}
where we only keep contributions which are consistent with the IA. Again within the IA, the static correlation function in Eq.~(\ref{eq:appendix-dynamic-structure-factor}) can be approximated by
\begin{equation}
\left\langle \hat{\psi}^{\dagger}(\mathbf{k}) \hat{\psi}(\mathbf{k}+\mathbf{q}) \right\rangle \left\langle \hat{\psi}^{\dagger}(\mathbf{k}^{\prime}) \hat{\psi}(\mathbf{k}^{\prime}-\mathbf{q}) \right\rangle +
\left\langle \hat{\psi}^{\dagger}(\mathbf{k}) \hat{\psi}(\mathbf{k}^{\prime}-\mathbf{q}) \right\rangle \left\langle \hat{\psi}(\mathbf{k}+\mathbf{q}) \hat{\psi}^{\dagger}(\mathbf{k}^{\prime}) \right\rangle.
\label{eq:appendix-correlation-functions-split}
\end{equation}
The correlation functions in the first term are negligible for large $q$, while those in the second term may be rewritten to give
\begin{eqnarray}
S(\mathbf{q},\omega) & = & \int \frac{d\mathbf{k}d\mathbf{k}^{\prime}}{(2\pi)^{3}} \delta (\omega - q^{2}/2m - \mathbf{q} \cdot \mathbf{k}/m ) \Bigg[\left\langle \hat{\psi}^{\dagger}(\mathbf{k}) \hat{\psi}(\mathbf{k}^{\prime}-\mathbf{q}) \right\rangle \delta(\mathbf{k}^{\prime}-\mathbf{k}-\mathbf{q}) \nonumber\\
 & & \qquad \qquad \qquad \qquad + \left\langle \hat{\psi}^{\dagger}(\mathbf{k}) \hat{\psi}(\mathbf{k}^{\prime}-\mathbf{q}) \right\rangle \left\langle \hat{\psi}^{\dagger}(\mathbf{k}^{\prime}) \hat{\psi}(\mathbf{k}+\mathbf{q}) \right\rangle \Bigg].
\label{eq:appendix-IA-dynamic-structure-factor}
\end{eqnarray}
Clearly for large $q \gg k,k^{\prime}$, the correlation functions in the second term in Eq.~(\ref{eq:appendix-IA-dynamic-structure-factor}) are negligible. Thus we finally arrive at the usual expression \cite{zambelli-pra-61-063608,griffin-excitations} for $S(\mathbf{q},\omega)$ in the IA, as given in Eq.~(\ref{eq:IA-dynamic-structure-factor}) of Section \ref{sec:Correlation-functions}.

\begin{acknowledgments}
AG would like to thank Alain Aspect and his group for discussions
about their Bragg scattering experiments \cite{gerbier-cond-mat-0210206}.
Both authors would like to thank NSERC of Canada for research support.
\end{acknowledgments}

\begin{thebibliography}{33}
\expandafter\ifx\csname natexlab\endcsname\relax\def\natexlab#1{#1}\fi
\expandafter\ifx\csname bibnamefont\endcsname\relax
  \def\bibnamefont#1{#1}\fi
\expandafter\ifx\csname bibfnamefont\endcsname\relax
  \def\bibfnamefont#1{#1}\fi
\expandafter\ifx\csname citenamefont\endcsname\relax
  \def\citenamefont#1{#1}\fi
\expandafter\ifx\csname url\endcsname\relax
  \def\url#1{\texttt{#1}}\fi
\expandafter\ifx\csname urlprefix\endcsname\relax\def\urlprefix{URL }\fi
\providecommand{\bibinfo}[2]{#2}
\providecommand{\eprint}[2][]{\url{#2}}

\bibitem[{\citenamefont{Popov}(1987)}]{popov-collective-excitations}
\bibinfo{author}{\bibfnamefont{V.~N.}~\bibnamefont{Popov}},
  \emph{\bibinfo{title}{Functional integrals and collective excitations}}
  (\bibinfo{publisher}{Cambridge University Press}, \bibinfo{address}{New
  York}, \bibinfo{year}{1987}).

\bibitem[{\citenamefont{G{\"o}rlitz et~al.}(2001)\citenamefont{G{\"o}rlitz,
  Vogels, Leanhardt, Raman, Gustavson, Abo-Shaeer, Chikkatur, Gupta, Inouye,
  Rosenband, and Ketterle}}]{gorlitz-prl-87-130402}
\bibinfo{author}{\bibfnamefont{A.}~\bibnamefont{G{\"o}rlitz}},
  \bibinfo{author}{\bibfnamefont{J.~M.}~\bibnamefont{Vogels}},
  \bibinfo{author}{\bibfnamefont{A.~E.}~\bibnamefont{Leanhardt}},
  \bibinfo{author}{\bibfnamefont{C.}~\bibnamefont{Raman}},
  \bibinfo{author}{\bibfnamefont{T.~L.}~\bibnamefont{Gustavson}},
  \bibinfo{author}{\bibfnamefont{J.~R.}~\bibnamefont{Abo-Shaeer}},
  \bibinfo{author}{\bibfnamefont{A.~P.}~\bibnamefont{Chikkatur}},
  \bibinfo{author}{\bibfnamefont{S.}~\bibnamefont{Gupta}},
  \bibinfo{author}{\bibfnamefont{S.}~\bibnamefont{Inouye}},
  \bibinfo{author}{\bibfnamefont{T.}~\bibnamefont{Rosenband}}, \bibnamefont{and}
  \bibinfo{author}{\bibfnamefont{W.}~\bibnamefont{Ketterle}},
  \bibinfo{journal}{Phys. Rev. Lett.}
  \textbf{\bibinfo{volume}{87}}, \bibinfo{pages}{130402}
  (\bibinfo{year}{2001}).

\bibitem[{\citenamefont{Ho and Ma}(1999)}]{ho-jltp-115-61}
\bibinfo{author}{\bibfnamefont{T.-L.} \bibnamefont{Ho}} \bibnamefont{and}
  \bibinfo{author}{\bibfnamefont{M.}~\bibnamefont{Ma}}, \bibinfo{journal}{J.
  Low Temp. Phys.} \textbf{\bibinfo{volume}{115}}, \bibinfo{pages}{61}
  (\bibinfo{year}{1999}).

\bibitem[{\citenamefont{Petrov et~al.}(2000{\natexlab{a}})\citenamefont{Petrov,
  Holzmann, and Shlyapnikov}}]{petrov-prl-84-2551}
\bibinfo{author}{\bibfnamefont{D.}~\bibnamefont{Petrov}},
  \bibinfo{author}{\bibfnamefont{M.}~\bibnamefont{Holzmann}}, \bibnamefont{and}
  \bibinfo{author}{\bibfnamefont{G.}~\bibnamefont{Shlyapnikov}},
  \bibinfo{journal}{Phys. Rev. Lett.} \textbf{\bibinfo{volume}{84}},
  \bibinfo{pages}{2551} (\bibinfo{year}{2000}{\natexlab{a}}).

\bibitem[{\citenamefont{Petrov et~al.}(2000{\natexlab{b}})\citenamefont{Petrov,
  Shlyapnikov, and Walraven}}]{petrov-prl-85-3745}
\bibinfo{author}{\bibfnamefont{D.~S.}~\bibnamefont{Petrov}},
  \bibinfo{author}{\bibfnamefont{G.~V.}~\bibnamefont{Shlyapnikov}},
  \bibnamefont{and} \bibinfo{author}{\bibfnamefont{J.~T.~M.}~\bibnamefont{Walraven}},
  \bibinfo{journal}{Phys. Rev. Lett.} \textbf{\bibinfo{volume}{85}},
  \bibinfo{pages}{3745} (\bibinfo{year}{2000}{\natexlab{b}}).

\bibitem[{\citenamefont{Petrov et~al.}(2001)\citenamefont{Petrov, Shlyapnikov,
  and Walraven}}]{petrov-prl-87-050404}
\bibinfo{author}{\bibfnamefont{D.~S.}~\bibnamefont{Petrov}},
  \bibinfo{author}{\bibfnamefont{G.~V.}~\bibnamefont{Shlyapnikov}},
  \bibnamefont{and} \bibinfo{author}{\bibfnamefont{J.~T.~M.}~\bibnamefont{Walraven}},
  \bibinfo{journal}{Phys. Rev. Lett.} \textbf{\bibinfo{volume}{87}},
  \bibinfo{pages}{050404} (\bibinfo{year}{2001}).

\bibitem[{\citenamefont{Shlyapnikov}(2001)}]{shlyapnikov-cargese-2-407}
\bibinfo{author}{\bibfnamefont{G.~V.}~\bibnamefont{Shlyapnikov}},
  \bibinfo{journal}{C. R. Acad. Sci. Paris, Ser. IV}
  \textbf{\bibinfo{volume}{2}}, \bibinfo{pages}{407} (\bibinfo{year}{2001}).

\bibitem[{\citenamefont{Dettmer et~al.}(2001)\citenamefont{Dettmer, Hellweg,
  Ryytty, Arlt, Ertmer, Sengstock, Petrov, Shlyapnikov, Kreutzmann, Santos, and 
  Lewenstein}}]{dettmer-prl-87-160406}
\bibinfo{author}{\bibfnamefont{S.}~\bibnamefont{Dettmer}},
  \bibinfo{author}{\bibfnamefont{D.}~\bibnamefont{Hellweg}},
  \bibinfo{author}{\bibfnamefont{P.}~\bibnamefont{Ryytty}},
  \bibinfo{author}{\bibfnamefont{J.~J.}~\bibnamefont{Arlt}},
  \bibinfo{author}{\bibfnamefont{W.}~\bibnamefont{Ertmer}},
  \bibinfo{author}{\bibfnamefont{K.}~\bibnamefont{Sengstock}},
  \bibinfo{author}{\bibfnamefont{D.~S.}~\bibnamefont{Petrov}},
  \bibinfo{author}{\bibfnamefont{G.~V.}~\bibnamefont{Shlyapnikov}},
  \bibinfo{author}{\bibfnamefont{H.}~\bibnamefont{Kreutzmann}},
  \bibinfo{author}{\bibfnamefont{L.}~\bibnamefont{Santos}}, \bibnamefont{and}
  \bibinfo{author}{\bibfnamefont{M.}~\bibnamefont{Lewenstein}},
  \bibinfo{journal}{Phys. Rev. Lett.}
  \textbf{\bibinfo{volume}{87}}, \bibinfo{pages}{160406}
  (\bibinfo{year}{2001}).

\bibitem[{\citenamefont{Gerbier et~al.}({\natexlab{a}})\citenamefont{Gerbier,
  Richard, Thywissen, Hugbert, Bouyer, and Aspect}}]{gerbier-cond-mat-0210206}
\bibinfo{author}{\bibfnamefont{F.}~\bibnamefont{Gerbier}},
  \bibinfo{author}{\bibfnamefont{S.}~\bibnamefont{Richard}},
  \bibinfo{author}{\bibfnamefont{J.~H.}~\bibnamefont{Thywissen}},
  \bibinfo{author}{\bibfnamefont{M.}~\bibnamefont{Hugbert}},
  \bibinfo{author}{\bibfnamefont{P.}~\bibnamefont{Bouyer}}, \bibnamefont{and}
  \bibinfo{author}{\bibfnamefont{A.}~\bibnamefont{Aspect}},
  \bibinfo{note}{cond-mat/0210206}.

\bibitem[{\citenamefont{Stenger et~al.}(1999)\citenamefont{Stenger, Inouye,
  Chikkatur, Stamper-Kurn, Pritchard, and Ketterle}}]{stenger-prl-82-4569}
\bibinfo{author}{\bibfnamefont{J.}~\bibnamefont{Stenger}},
  \bibinfo{author}{\bibfnamefont{S.}~\bibnamefont{Inouye}},
  \bibinfo{author}{\bibfnamefont{A.~P.}~\bibnamefont{Chikkatur}},
  \bibinfo{author}{\bibfnamefont{D.~M.}~\bibnamefont{Stamper-Kurn}},
  \bibinfo{author}{\bibfnamefont{D.~E.}~\bibnamefont{Pritchard}},
  \bibnamefont{and} \bibinfo{author}{\bibfnamefont{W.}~\bibnamefont{Ketterle}},
  \bibinfo{journal}{Phys. Rev. Lett.} \textbf{\bibinfo{volume}{82}},
  \bibinfo{pages}{4569} (\bibinfo{year}{1999}).

\bibitem[{\citenamefont{Stamper-Kurn et~al.}(1999)\citenamefont{Stamper-Kurn,
  Chikkatur, G{\"o}rlitz, Inouye, Gupta, Pritchard, and
  Ketterle}}]{stamper-kurn-prl-83-2876}
\bibinfo{author}{\bibfnamefont{D.~M.}~\bibnamefont{Stamper-Kurn}},
  \bibinfo{author}{\bibfnamefont{A.~P.}~\bibnamefont{Chikkatur}},
  \bibinfo{author}{\bibfnamefont{A.}~\bibnamefont{G{\"o}rlitz}},
  \bibinfo{author}{\bibfnamefont{S.}~\bibnamefont{Inouye}},
  \bibinfo{author}{\bibfnamefont{S.}~\bibnamefont{Gupta}},
  \bibinfo{author}{\bibfnamefont{D.~E.}~\bibnamefont{Pritchard}},
  \bibnamefont{and} \bibinfo{author}{\bibfnamefont{W.}~\bibnamefont{Ketterle}},
  \bibinfo{journal}{Phys. Rev. Lett.} \textbf{\bibinfo{volume}{83}},
  \bibinfo{pages}{2876} (\bibinfo{year}{1999}).

\bibitem[{\citenamefont{Ketterle}(1999)}]{ketterle-les-houches-1999}
\bibinfo{author}{\bibfnamefont{W.}~\bibnamefont{Ketterle}}, in
  \emph{\bibinfo{booktitle}{Coherent {A}tomic {M}atter {W}aves}}, edited by
  \bibinfo{editor}{\bibfnamefont{C.}~\bibnamefont{Westbrook}} \bibnamefont{and}
  \bibinfo{editor}{\bibfnamefont{F.}~\bibnamefont{David}}
  (\bibinfo{publisher}{Springer-Verlag}, \bibinfo{address}{New York},
  \bibinfo{year}{1999}), Proceedings of {L}es {H}ouches 1999 {S}ummer {S}chool,
  \emph{{S}ession {LXXII}}.

\bibitem[{\citenamefont{Vogels et~al.}(2002)\citenamefont{Vogels, Xu, Raman,
  Abo-Shaeer, and Ketterle}}]{vogels-prl-88-060402}
\bibinfo{author}{\bibfnamefont{J.~M.}~\bibnamefont{Vogels}},
  \bibinfo{author}{\bibfnamefont{K.}~\bibnamefont{Xu}},
  \bibinfo{author}{\bibfnamefont{C.}~\bibnamefont{Raman}},
  \bibinfo{author}{\bibfnamefont{J.~R.}~\bibnamefont{Abo-Shaeer}},
  \bibnamefont{and} \bibinfo{author}{\bibfnamefont{W.}~\bibnamefont{Ketterle}},
  \bibinfo{journal}{Phys. Rev. Lett.} \textbf{\bibinfo{volume}{88}},
  \bibinfo{pages}{060402} (\bibinfo{year}{2002}).

\bibitem[{\citenamefont{Steinhauer et~al.}(2002)\citenamefont{Steinhauer,
  Ozeri, Katz, and Davidson}}]{steinhauer-prl-88-120407}
\bibinfo{author}{\bibfnamefont{J.}~\bibnamefont{Steinhauer}},
  \bibinfo{author}{\bibfnamefont{R.}~\bibnamefont{Ozeri}},
  \bibinfo{author}{\bibfnamefont{N.}~\bibnamefont{Katz}}, \bibnamefont{and}
  \bibinfo{author}{\bibfnamefont{N.}~\bibnamefont{Davidson}},
  \bibinfo{journal}{Phys. Rev. Lett.} \textbf{\bibinfo{volume}{88}},
  \bibinfo{pages}{120407} (\bibinfo{year}{2002}).

\bibitem[{\citenamefont{Japha et~al.}(1999)\citenamefont{Japha, Choi, Burnett,
  and Band}}]{japha-prl-82-1079}
\bibinfo{author}{\bibfnamefont{Y.}~\bibnamefont{Japha}},
  \bibinfo{author}{\bibfnamefont{S.}~\bibnamefont{Choi}},
  \bibinfo{author}{\bibfnamefont{K.}~\bibnamefont{Burnett}}, \bibnamefont{and}
  \bibinfo{author}{\bibfnamefont{Y.~B.}~\bibnamefont{Band}},
  \bibinfo{journal}{Phys. Rev. Lett.} \textbf{\bibinfo{volume}{82}},
  \bibinfo{pages}{1079} (\bibinfo{year}{1999}).

\bibitem[{\citenamefont{Choi et~al.}(2000)\citenamefont{Choi, Japha, and
  Burnett}}]{choi-pra-61-063606}
\bibinfo{author}{\bibfnamefont{S.}~\bibnamefont{Choi}},
  \bibinfo{author}{\bibfnamefont{Y.}~\bibnamefont{Japha}}, \bibnamefont{and}
  \bibinfo{author}{\bibfnamefont{K.}~\bibnamefont{Burnett}},
  \bibinfo{journal}{Phys. Rev. A} \textbf{\bibinfo{volume}{61}},
  \bibinfo{pages}{063606} (\bibinfo{year}{2000}).

\bibitem[{\citenamefont{Luxat and Griffin}(2002)}]{luxat-pra-65-043618}
\bibinfo{author}{\bibfnamefont{D.~L.}~\bibnamefont{Luxat}} \bibnamefont{and}
  \bibinfo{author}{\bibfnamefont{A.}~\bibnamefont{Griffin}},
  \bibinfo{journal}{Phys. Rev. A} \textbf{\bibinfo{volume}{65}},
  \bibinfo{pages}{043618} (\bibinfo{year}{2002}).

\bibitem[{\citenamefont{Yip}(2001)}]{yip-prl-87-130401}
\bibinfo{author}{\bibfnamefont{S.~K.}~\bibnamefont{Yip}}, \bibinfo{journal}{Phys.
  Rev. Lett.} \textbf{\bibinfo{volume}{87}}, \bibinfo{pages}{130401}
  (\bibinfo{year}{2001}).

\bibitem[{\citenamefont{Kang et~al.}(2000)\citenamefont{Kang, Stormer,
  Pfeiffer, Baldwin, and West}}]{kang-nature-403-59}
\bibinfo{author}{\bibfnamefont{W.}~\bibnamefont{Kang}},
  \bibinfo{author}{\bibfnamefont{H.~L.}~\bibnamefont{Stormer}},
  \bibinfo{author}{\bibfnamefont{L.~N.}~\bibnamefont{Pfeiffer}},
  \bibinfo{author}{\bibfnamefont{K.~W.}~\bibnamefont{Baldwin}}, \bibnamefont{and}
  \bibinfo{author}{\bibfnamefont{K.~W.}~\bibnamefont{West}},
  \bibinfo{journal}{Nature} \textbf{\bibinfo{volume}{403}}, \bibinfo{pages}{59}
  (\bibinfo{year}{2000}).

\bibitem[{\citenamefont{Auslaender et~al.}(2002)\citenamefont{Auslaender,
  Yacoby, de~Picciotto, Baldwin, Pfeiffer, and
  West}}]{auslaender-science-295-825}
\bibinfo{author}{\bibfnamefont{O.~M.}~\bibnamefont{Auslaender}},
  \bibinfo{author}{\bibfnamefont{A.}~\bibnamefont{Yacoby}},
  \bibinfo{author}{\bibfnamefont{R.}~\bibnamefont{de~Picciotto}},
  \bibinfo{author}{\bibfnamefont{K.~W.}~\bibnamefont{Baldwin}},
  \bibinfo{author}{\bibfnamefont{L.~N.}~\bibnamefont{Pfeiffer}}, \bibnamefont{and}
  \bibinfo{author}{\bibfnamefont{K.~W.}~\bibnamefont{West}},
  \bibinfo{journal}{Science} \textbf{\bibinfo{volume}{295}},
  \bibinfo{pages}{825} (\bibinfo{year}{2002}).

\bibitem[{\citenamefont{Menotti and Stringari}(2002)}]{menotti-pra-66-043610}
\bibinfo{author}{\bibfnamefont{C.}~\bibnamefont{Menotti}} \bibnamefont{and}
  \bibinfo{author}{\bibfnamefont{S.}~\bibnamefont{Stringari}},
  \bibinfo{journal}{Phys. Rev. A} \textbf{\bibinfo{volume}{66}},
  \bibinfo{pages}{043610} (\bibinfo{year}{2002}).

\bibitem[{\citenamefont{Stringari}(1996)}]{stringari-prl-77-2360}
\bibinfo{author}{\bibfnamefont{S.}~\bibnamefont{Stringari}},
  \bibinfo{journal}{Phys. Rev. Lett.} \textbf{\bibinfo{volume}{77}},
  \bibinfo{pages}{2360} (\bibinfo{year}{1996}).

\bibitem[{\citenamefont{Wu and Griffin}(1996)}]{wu-pra-54-4204}
\bibinfo{author}{\bibfnamefont{W.-C.} \bibnamefont{Wu}} \bibnamefont{and}
  \bibinfo{author}{\bibfnamefont{A.}~\bibnamefont{Griffin}},
  \bibinfo{journal}{Phys. Rev. A} \textbf{\bibinfo{volume}{54}},
  \bibinfo{pages}{4204} (\bibinfo{year}{1996}).

\bibitem[{\citenamefont{Zambelli et~al.}(2000)\citenamefont{Zambelli,
  Pitaevskii, Stamper-Kurn, and Stringari}}]{zambelli-pra-61-063608}
\bibinfo{author}{\bibfnamefont{F.}~\bibnamefont{Zambelli}},
  \bibinfo{author}{\bibfnamefont{L.}~\bibnamefont{Pitaevskii}},
  \bibinfo{author}{\bibfnamefont{D.~M.}~\bibnamefont{Stamper-Kurn}},
  \bibnamefont{and}
  \bibinfo{author}{\bibfnamefont{S.}~\bibnamefont{Stringari}},
  \bibinfo{journal}{Phys. Rev. A} \textbf{\bibinfo{volume}{61}},
  \bibinfo{pages}{063608} (\bibinfo{year}{2000}).

\bibitem[{\citenamefont{Olshanii}(1998)}]{olshanii-prl-81-938}
\bibinfo{author}{\bibfnamefont{M.}~\bibnamefont{Olshanii}},
  \bibinfo{journal}{Phys. Rev. Lett.} \textbf{\bibinfo{volume}{81}},
  \bibinfo{pages}{938} (\bibinfo{year}{1998}).

\bibitem[{\citenamefont{Haldane}(1981)}]{haldane-prl-47-1840}
\bibinfo{author}{\bibfnamefont{F.~D.~M.}~\bibnamefont{Haldane}},
  \bibinfo{journal}{Phys. Rev. Lett.} \textbf{\bibinfo{volume}{47}},
  \bibinfo{pages}{1840} (\bibinfo{year}{1981}).

\bibitem[{\citenamefont{Gangardt and Shlyapnikov}()}]{gangardt-lanl-0207338}
\bibinfo{author}{\bibfnamefont{D.~M.}~\bibnamefont{Gangardt}} \bibnamefont{and}
  \bibinfo{author}{\bibfnamefont{G.~V.}~\bibnamefont{Shlyapnikov}},
  \bibinfo{note}{cond-mat/0207338}.

\bibitem[{\citenamefont{Stringari}(1998)}]{stringari-pra-58-2385}
\bibinfo{author}{\bibfnamefont{S.}~\bibnamefont{Stringari}},
  \bibinfo{journal}{Phys. Rev. A} \textbf{\bibinfo{volume}{58}},
  \bibinfo{pages}{2385} (\bibinfo{year}{1998}).

\bibitem[{\citenamefont{{Al Khawaja} et~al.}(2002)\citenamefont{{Al Khawaja},
  Andersen, Proukakis, and Stoof}}]{khawaja-pra-66-013615}
\bibinfo{author}{\bibfnamefont{U.}~\bibnamefont{{Al Khawaja}}},
  \bibinfo{author}{\bibfnamefont{J.~O.}~\bibnamefont{Andersen}},
  \bibinfo{author}{\bibfnamefont{N.~P.}~\bibnamefont{Proukakis}},
  \bibnamefont{and} \bibinfo{author}{\bibfnamefont{H.~T.~C.}~\bibnamefont{Stoof}},
  \bibinfo{journal}{Phys. Rev. A} \textbf{\bibinfo{volume}{66}},
  \bibinfo{pages}{013615} (\bibinfo{year}{2002}).

\bibitem[{\citenamefont{Andersen et~al.}(2002)\citenamefont{Andersen, {Al
  Khawaja}, and Stoof}}]{andersen-prl-88-070407}
\bibinfo{author}{\bibfnamefont{J.~O.}~\bibnamefont{Andersen}},
  \bibinfo{author}{\bibfnamefont{U.}~\bibnamefont{{Al Khawaja}}},
  \bibnamefont{and} \bibinfo{author}{\bibfnamefont{H.~T.~C.}~\bibnamefont{Stoof}},
  \bibinfo{journal}{Phys. Rev. Lett.} \textbf{\bibinfo{volume}{88}},
  \bibinfo{pages}{070407} (\bibinfo{year}{2002}).

\bibitem[{\citenamefont{Gerbier et~al.}({\natexlab{b}})\citenamefont{Gerbier,
  Thywissen, Richard, Hugbart, Bouyer, and Aspect}}]{gerbier-cond-mat-0211094}
\bibinfo{author}{\bibfnamefont{F.}~\bibnamefont{Gerbier}},
  \bibinfo{author}{\bibfnamefont{J.~H.}~\bibnamefont{Thywissen}},
  \bibinfo{author}{\bibfnamefont{S.}~\bibnamefont{Richard}},
  \bibinfo{author}{\bibfnamefont{M.}~\bibnamefont{Hugbart}},
  \bibinfo{author}{\bibfnamefont{P.}~\bibnamefont{Bouyer}}, \bibnamefont{and}
  \bibinfo{author}{\bibfnamefont{A.}~\bibnamefont{Aspect}},
  \bibinfo{note}{cond-mat/0211094}.

\bibitem[{\citenamefont{Weling et~al.}(1983)\citenamefont{Weling, Griffin, and
  Carrington}}]{weling-prb-28-5296}
\bibinfo{author}{\bibfnamefont{F.}~\bibnamefont{Weling}},
  \bibinfo{author}{\bibfnamefont{A.}~\bibnamefont{Griffin}}, \bibnamefont{and}
  \bibinfo{author}{\bibfnamefont{M.}~\bibnamefont{Carrington}},
  \bibinfo{journal}{Phys. Rev. B} \textbf{\bibinfo{volume}{28}},
  \bibinfo{pages}{5296} (\bibinfo{year}{1983}).

\bibitem[{\citenamefont{Griffin}(1993)}]{griffin-excitations}
\bibinfo{author}{\bibfnamefont{A.}~\bibnamefont{Griffin}},
  \emph{\bibinfo{title}{Excitations in a Bose-Condensed Liquid}}
  (\bibinfo{publisher}{Cambridge University Press}, \bibinfo{address}{New
  York}, \bibinfo{year}{1993}).

\end{thebibliography}

\end{document}